\documentclass[%
reprint,
 amsmath,amssymb,
 aps,
pra,
]{revtex4-1}
\usepackage{amsmath}

\usepackage{graphicx}% Include figure files
\usepackage{subfigure}
\usepackage{dcolumn}% Align table columns on decimal point
\usepackage{bm}
\usepackage{epstopdf}
\usepackage{epsfig}% bold math
\usepackage{color}
\usepackage[colorlinks,
            linkcolor=blue,
            anchorcolor=blue,
            citecolor=blue,
            urlcolor=blue
            ]{hyperref}

\begin{document}

%\preprint{APS/123-QED}

\title{Exponential Sensitivity Revival of Noisy non-Hermitian Quantum Sensing with Two-photon Drives}% Force line breaks with \\
\author{Liying Bao$^{1,2,3}$, Bo Qi$^{1,2\star}$, Franco Nori$^{4,5,6}$, Daoyi Dong$^{7}$
\\
$^{1}$\textit{Key Laboratory of Systems and Control, Academy of Mathematics and Systems Science, Chinese Academy of Sciences, Beijing 100190, China}\\
$^{2}$\textit{University of Chinese Academy of Sciences, Beijing 100049, China}\\
$^{3}$\textit{Civil Aviation University of China, Tianjin 300300, China}\\
$^{4}$\textit{Theoretical Quantum Physics Laboratory, RIKEN, Saitama, 351-0198, Japan}\\
$^{5}$\textit{Quantum Computing Center, RIKEN, Saitama, 351-0198, Japan}\\
$^{6}$\textit{Physics Department, The University of Michigan, Ann Arbor, Michigan 48109, USA}\\
$^{7}$\textit{School of Engineering, Australian National University, Canberra ACT 2601, Australia}\\
$^\star$\textit{qibo@amss.ac.cn}}

\date{\today}% It is always \today, today,

\begin{abstract}
Unique properties of multimode non-Hermitian lattice dynamics can be utilized to construct exponentially sensitive sensors. However, the impact of noise remains unclear, which may severely degrade their sensitivity. We analytically characterize and highlight the impact of loss and gain on the sensitivity revival and stability of non-Hermitian sensors.  Defying the general belief that the superiority of quantum sensing will vanish in the presence of loss, we find that by proactively tuning the loss, the exponential sensitivity can be surprisingly regained when the sensing dynamics is stable. Furthermore, we prove that  gain is crucial to fully revive the ideally exponential sensitivity and to ensure the stability of non-Hermitian sensing by making a balanced loss and gain. Our work opens a new way to significantly enhance the sensitivity by proactively tuning the loss and gain, which may promote future quantum sensing and quantum engineering.

\end{abstract}

%\pacs{Valid PACS appear here} % PACS, the Physics and Astronomy
\maketitle

%\tableofcontents

\section{Introduction}
High precision sensors are ubiquitous and vitally important in both science and technology. Due to the high susceptibility of  the complex energy spectra of non-Hermitian (NH) Hamiltonians in response to small perturbations, NH sensors have been attracting increasing attention. Various unconventional properties  of NH systems have been studied to theoretically propose high sensitive sensors \cite{Peng2014,Nikzamir2024,ZhangJing2018,Ortolano2023,Lei2023,G.-Q.Zhang2021,El-Ganainy2018,Bliokh2019,Chu2020,Liu2019,Bensa2021,Lee2023,Okuma2020,Scheibner2020,Kawabata2020}, and some  architectures  have already been experimentally demonstrated\cite{Park2021,Chen2021,xu2022}. In this work, we investigate the sensitivity revival and stability of NH quantum sensing in noisy environments.

The pursuit of high sensitivity  is a fundamental objective in developing sensing technology. Recent progress has shown that the intriguing degeneracy property of NH systems can be employed to enhance the sensitivity of sensors operating at finely tuned exceptional points (EPs), where the coalesced eigenenergies have a diverging susceptibility to small perturbations \cite{Wiersig2014,ozdemir2019, Zhang2019, Wiersig2016, Chen2019, WangGao2020, Jiang2022, Pap2021}.  However, to assess the performance of sensors based on EPs,  we should also take into account of the effect of the coalesced eigenstates, which may counteract the  diverging susceptibility of eigenenergies \cite{Wiersig2016, Chen2019}. Other distinct properties of NH systems have also been harnessed to enhance the sensitivity of  NH sensors, which do not necessarily work at EPs. As studied in  \cite{Lau2018},  nonreciprocity \cite{Tzuang2014,TangJ2022, Tang2022,Sounas2017, Lau2018, Bao2021} can be identified as a powerful resource for sensing,  since it allows one to exceed the fundamental bounds constraining conventional, reciprocal sensors~\cite{Lau2018}.  Remarkably, a class of sensors having exponential sensitivity have been theoretically proposed~\cite{McDonald2020,Budich2020,Qin2018,Koch2021,Baoli2021}. The drastic enhancements rely upon the strikingly anomalous sensitivity to the boundary conditions of NH systems. Furthermore, the implication of optimizing controllable parameters in attaining an exponential enhancement was investigated in \cite{Baoli2021,Chen2021}.

In practical applications the existence of noise is unavoidable, which may severely degrade  the performance, such as the stability and  sensitivity of NH sensors.  Since most sensing schemes are measured at equilibrium states, a {\it stable} sensing dynamics is a fundamental requirement for achieving these high precision sensing. 
{\it Loss} noise has an essential impact on the attainable sensitivity in quantum sensing. For conventional sensors, it is well-known that by using quantum strategies, the precision can be scaled as $1/N$ in terms of the number $N$ of quantum resources for noiseless processes\cite{Giovannetti2006,Napolitano2011,Thomas2011,Hou2019,Hou2020,Yuan2015}. However, it has been demonstrated in  \cite{Escher2011, Datta2011,Demkowicz2012,Zhou2018} that even a weak loss noise can quickly degrade the precision from $1/N$  to $1/\sqrt{N}$, independently of the initial state of the probes and even regardless of the use of adaptive feedback.  This is very frustrating and it has been a general belief that advantages of quantum sensing will soon vanish in the presence of loss. {\it Gain} has been demonstrated to be a necessary ingredient to have an enhanced signal power in NH sensing \cite{Lau2018}, whereas  too much gain may result in an unstable sensing dynamics. 

Since manipulating loss and gain has become feasible~\cite{Feng2014,Liuyl2017,Ren2022,Abbas2023,Xue20232}, it is of vital importance to understand the impact of loss and gain on the sensitivity and stability of NH sensing. Different from existing results, in this work, our aim is to find out whether we can achieve {\it exponentially enhanced and stable} NH quantum sensing  by {\it proactively} tuning the loss and gain.  In general, the loss and gain  bring about two effects for the NH sensing dynamics: one is the diffusion noise that may be further amplified during the sensing and then severely degrade the sensitivity; the other one is the dissipative drift that may lead to system instability. 

We find conditions to fully recover the ideally noiseless sensitivity for noisy NH quantum sensing in both the perturbation regime and the case beyond linear response. To be specific, we discover that the coupling  of  loss plays a pivotal role in obtaining exponential sensitivity revival in our setting.  Counterintuitively, we find that by proactively tuning the loss couplings properly, an exponential signal-to-noise ratio (SNR)  can be surprisingly regained when the sensing dynamics is stable. We  further point out that balanced gain and loss is vital to fully recover the ideally noiseless sensitivity and to ensure the stability of the NH sensing dynamics. We also analyze the robustness of the sensitivity under the designed loss and gain and provide a guideline on how to realize our strategy in practice through concrete examples.

\section{Setup of noisy NH sensors}
\begin{figure}[!h]
\centering
\includegraphics[scale=0.58]{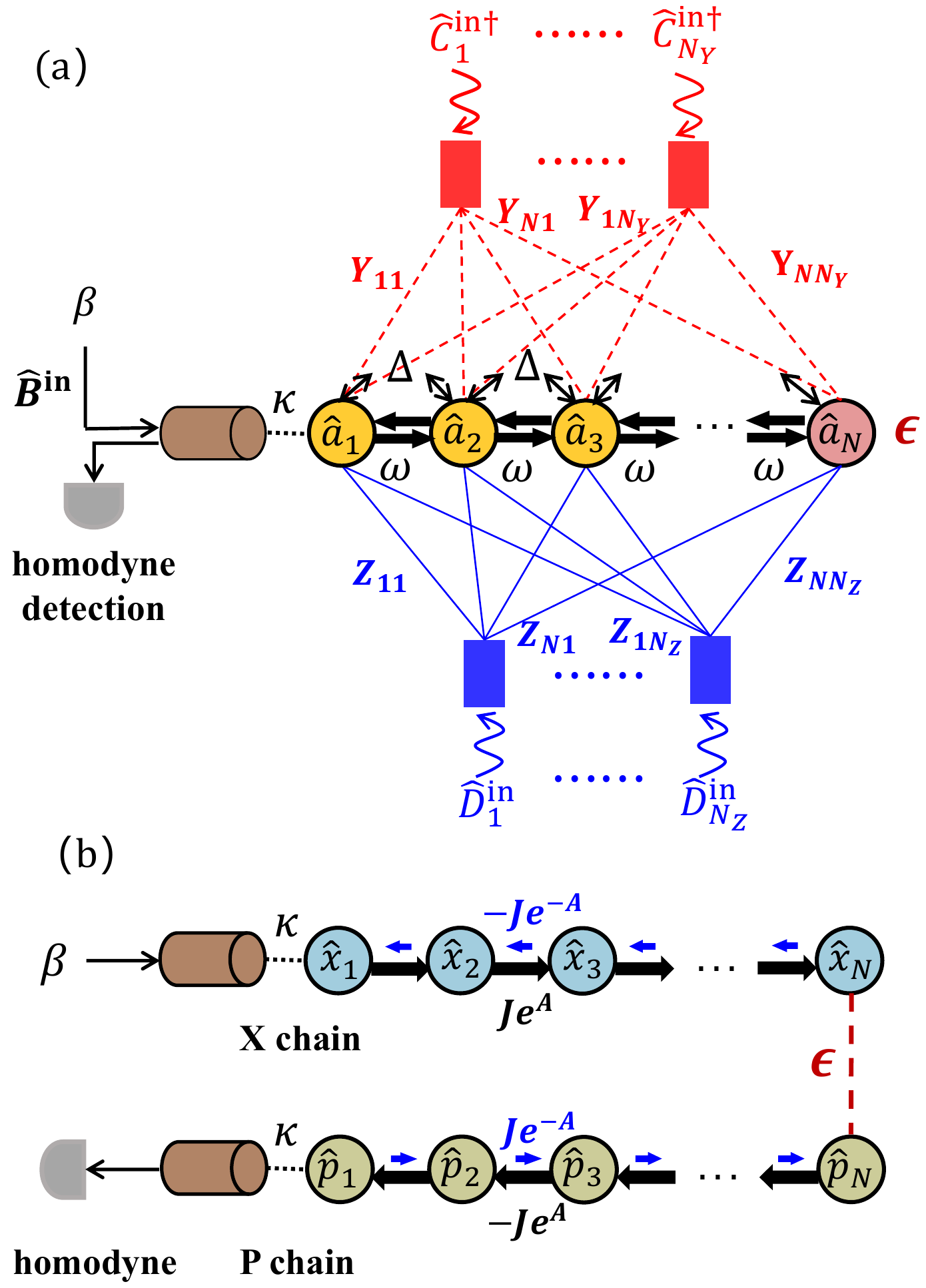}
%\vspace{-0.5em}
\caption{(Color online) A general multimode noisy NH setup.  (a) The setup consists of a 1-dimensional chain of $N$  bosonic modes. The parameter to be detected is $\epsilon$, which represents a small change in the resonance frequency of the last site. To detect $\epsilon$, a coherent drive $\beta$ accompanied by  quantum noise $\hat{B}^{\textsf{in}}$ is injected into the chain at mode 1 through an input-output waveguide with coupling rate $\kappa$. The reflected field  is measured by homodyne detection.  The modes are coupled via nearest neighbour hopping $w$ and coherent two-photon drive $\Delta$. To account for the noise, couplings between the modes and the loss/gain baths (blue solid/red dashed) are included.  The coupling rate between the $i$th mode and the $j$th loss (gain) bath is described by $Z_{ij}$  ($Y_{ij}$). (b) The nonreciprocal amplification between modes can be described by  two $N$-site NH Hatano-Nelson chains \cite{Hatano1997,Hatano1996,Leefmans2024} with effective hopping amplitude $J$ and amplification factor $A$. For the top (bottom) \textrm{X} (\textrm{P}) chain, hopping to the right is a factor of $e^{2A}$ larger (smaller) than hopping to the left. The last modes of the two chains are coupled due to the presence of small tunneling with amplitude $\epsilon$, allowing the signal to be transmitted between the two chains.}
\end{figure}

In the absence of noise, an exponentially enhanced quantum sensing scheme was proposed in~\cite{McDonald2020} based on NH lattice dynamics. In this work, we adopt the model in~\cite{McDonald2020}, and then further investigate the impact of loss and gain on the sensitivity and stability of NH sensing.

A generic multimode noisy NH  setup is illustrated in Fig.~1(a). Consider a 1-dimensional array of $N$ bosonic modes, and let  $\hat{a}_i$ denote the mode annihilation operator on the $i$th site. Our aim is to detect a small perturbation $\epsilon$ of a perturbation Hamiltonian $\epsilon \hat{V}$,  where $\hat{V}$ is a system operator. In Fig.~1(a) and Fig.~1(b),   $\hat{V}=\hat{a}_N^\dag\hat{a}_N$,  and thus the aim in this case is to estimate a small change $\epsilon$ in the resonance frequency of the last site. A general measurement strategy is to couple mode 1 to an input-output waveguide with rate $\kappa$, and then inject a coherent drive with amplitude $\beta$ at the resonant frequency of the mode. The reflected signal is measured by a homodyne detection \cite{Moiseyev2011} to infer $\epsilon$ \cite{Lau2018,Bao2021,Baoli2021,McDonald2020}.
In the rotating frame set by the mode resonance frequency, the system Hamiltonian reads 
\begin{flalign}
\hat{H}_S=\sum^{N-1}_{n=1}(i w \hat{a}^\dagger_{n+1}\hat{a}_n+i \Delta \hat{a}^\dagger_{n+1}\hat{a}^\dagger_n+\textit{h.c.}), 
\end{flalign} 
where  $\omega$ depicts the hopping between neighbor modes and $\Delta$ describes the nearest-neighbour two-photon drive \cite{McDonald2020,Scully1997}. We assume that $w>\Delta>0$. Up to now, this is the ideal model utilized in~\cite{McDonald2020}.

To fully account for the  noise, we couple the modes to $N_Z$ loss  and $N_Y$ gain baths, which are mutually independent.  Without loss of generality, the coupling rates are described  by the real matrices $Z$ and $Y$, respectively. The element $Z_{ij}$ ($Y_{ij}$) of the loss (gain) coupling matrix $Z$ ($Y$) describes the coupling rate between the $i$th mode and the $j$th loss (gain) bath.
Using the standard input-output theory \cite{Clerk2010}, the total effective Hamiltonian (see Appendix~\ref{appendixA} for details) reads 
\begin{flalign} \hat{H}[\epsilon]\!=\!\hat{H}_S\!+\!\epsilon\hat{V}\!+\!\hat{H}_\kappa\!+\!\hat{H}_{G}\!+\!\hat{H}_{L}\!-i\sqrt{\kappa}(\hat{a}_1^\dagger \beta\!-\!h.c.),
\end{flalign}
where $\hat{H}_{\kappa}$ describes the damping of mode 1 due to the coupling with the  waveguide,  while $\hat{H}_{G}$ and $\hat{H}_L$ describe the damping owing to the coupling with the gain  and loss baths, respectively. The Heisenberg-Langevin equations (see Appendix~\ref{appendixA} for details) read 
\begin{flalign}\label{anheisenberg}
\begin{aligned}
\dot{\hat{a}}_n=&w \hat{a}_{n-1}\!-\!w\hat{a}_{n+1}\!+\!\Delta\hat{a}^\dagger_{n+1}\!+\!\Delta\hat{a}^\dagger_{n-1}\!-\!i\epsilon[\hat{a}_n,\hat{V}]\\
&+\sum_{j=1}^{N_{Y}}\sum_{i=1}^{N}Y_{n,j}Y_{i,j}\hat{a}_i-\sum_{j=1}^{N_{Z}}\sum_{i=1}^{N}Z_{n,j}Z_{i,j}\hat{a}_i\\
&-\frac{\kappa}{2}\hat{a}_1\delta_{n,1}-\sqrt{\kappa}(\hat{B}^{\textsf{in}}+\beta)\delta_{n,1}\\
&-\sqrt{2}\Big{(}\sum_{j=1}^{N_{Y}}Y_{n,j}\hat{C}_{j}^{\textsf{in}\dagger}+\sum_{j=1}^{N_{Z}}Z_{n,j}\hat{D}_{j}^{\textsf{in}}\Big{)}.
\end{aligned}
\end{flalign}
Here, $\hat{B}^{\textsf{in}}$ denotes the quantum noise entering from the waveguide, and $\hat{C}_{j}^{\textsf{in}}$ ($\hat{D}_{j}^{\textsf{in}}$) are quantum noises arising from the gain (loss) process of the baths. To ensure  Markovian dynamics, $\hat{B}^{\textsf{in}}$, $\hat{C}_{j}^{\textsf{in}}$ and $\hat{D}_{j}^{\textsf{in}}$ are assumed to be quantum Gaussian white noise satisfying: 
\begin{flalign}
\begin{aligned}
\langle Q(t)Q^\dagger(t')\rangle&=(\bar{n}^{\textsf{th}}_Q+1)\delta(t-t'), \\
\langle Q^\dagger(t)Q(t')\rangle&=\bar{n}^{\textsf{th}}_Q\delta(t-t'),\\ \langle Q(t)Q(t')\rangle&=0, 
\end{aligned}
\end{flalign}
where $Q\in \{ \hat{B}^{\textsf{in}},~\hat{C}^{\textsf{in}}_j,~\hat{D}^{\textsf{in}}_j \}$, and there are no correlations between different noise operators. Here, $\bar{n}_{\textsf{th}}$ is the number of thermal quanta in the input field. Hereafter, we focus on the vacuum noise, namely, $\bar{n}^{\textsf{th}}_Q=0$.

To see clearly how the signal is amplified, we turn to the picture of canonical quadratures $\hat{x}_n$ and $\hat{p}_n$ related with $\hat{a}_n$ via $\hat{a}_n=(\hat{x}_n+i\hat{p}_n)/\sqrt{2}$. Define quadrature vectors $\hat{\mathbf{X}}=(\hat{x}_1,\hat{x}_2,\ldots,\hat{x}_N)^\top$ and $\hat{\mathbf{P}}=(\hat{p}_1,\hat{p}_2,\ldots,\hat{p}_N)^\top$, respectively. Then the Heisenberg-Langevin equations (see Appendix~\ref{appendixA} for details) turn to
\begin{equation}\label{main}
\begin{aligned}
\begin{aligned} 
 \begin{pmatrix}
 \dot{ \hat{\mathbf{X}} } \\
  \dot{\hat{\mathbf{P}}}  \\
 \end{pmatrix}=&\begin{pmatrix}
  h^\mathbb{X}\!+\!YY^\top\!-\!ZZ^\top &0 \\
  0 & h^\mathbb{P}\!+\!YY^\top\!-\!ZZ^\top  \\
 \end{pmatrix}\begin{pmatrix}
  \hat{\mathbf{X}}  \\
  \hat{\mathbf{P}}  \\
 \end{pmatrix}\\
 &-i\epsilon\begin{pmatrix}
 [\hat{\mathbf{X}},\hat{V}]\\
 [\hat{\mathbf{P}},\hat{V}]
 \end{pmatrix}-\vec{\beta}-\hat{\Omega}^{\textsf{in}}.
\end{aligned} \end{aligned}
\end{equation}
Here, $h^\mathbb{X}$ and $h^\mathbb{P}$ represent the ideally noiseless  dynamical matrices of the quadratures $\hat{\mathbf{X}}$ and $\hat{\mathbf{P}}$, respectively, which read
\begin{flalign}
\begin{aligned}
h^\mathbb{X}\!&=\!-\frac{\kappa}{2}|1\rangle\langle1|+\!\sum^{N-1}_{n=1}\!\Big(Je^{A}|n+1\rangle\langle n|\!-\!Je^{-A}|n\rangle\langle n+1|\Big),\\
h^\mathbb{P}\!&=\!-\frac{\kappa}{2}|1\rangle\langle1|+\!\sum^{N-1}_{n=1}\!\Big(Je^{-A}|n+1\rangle\langle n|\!-\!Je^{A}|n\rangle\langle n+1|\Big),
\end{aligned}
\end{flalign}
where $J= \sqrt{w^2-\Delta^2}$ denotes the hopping amplitude and the amplification factor
 $A$ is defined via 
\begin{flalign}
e^{2A}= \frac{w+\Delta}{w-\Delta}.
\end{flalign}
Due to  $h^\mathbb{X}$ and $h^\mathbb{P}$, we can find that for the  top \textrm{X} (bottom \textrm{P}) chain in Fig.~1(b),  hopping to the right is a factor of $e^{2A}$ larger (smaller) than hopping to the left.  The commutators with the perturbation Hamiltonian $\hat{V}$ are defined in an element-wise way, e.g.,  
\begin{flalign}
[\hat{\mathbf{X}},\hat{V}]=([\hat{x}_1, \hat{V}],\cdots,[\hat{x}_N, \hat{V}])^{\top}.
\end{flalign}
The coherent input vector
\begin{flalign}
\vec{\beta}=(\sqrt{2\kappa}\beta,0,0,\ldots,0)^\top, 
\end{flalign}
and  $\hat{\Omega}^{\textsf{in}}$ denotes the quantum noise vector (see Appendix~\ref{appendixA} for details), whose elements are described by
\begin{flalign}
\begin{aligned}
\hat{\Omega}^{\textsf{in}}_{i}&\!=\!\sqrt{\kappa}\hat{X}^{\textsf{in}}\delta_{i\!,\!1}\!+\!\sqrt{2}\Big{(}\sum_{j=1}^{N_Y}Y_{i\!,\!j}\hat{C}_{j,X}^{\textsf{in}}\!+\!\!\sum_{j=1}^{N_Z}\!Z_{i\!,\!j}\hat{D}_{j,X}^{\textsf{in}}\Big{)},\\
\hat{\Omega}^{\textsf{in}}_{i\!+\!N}&\!=\!\sqrt{\kappa}\hat{P}^{\textsf{in}}\delta_{i\!,\!1}\!+\!\sqrt{2}\Big{(}-\!\!\sum_{j=1}^{N_Y}Y_{i\!,\!j}\hat{C}_{j,P}^{\textsf{in}}\!+\!\!\sum_{j=1}^{N_Z}\!Z_{i\!,\!j}\hat{D}_{j,P}^{\textsf{in}}\Big{)},
\end{aligned}
\end{flalign}
for $i\in\{1,2,\ldots,N\}$, with $\hat{B}^{\textsf{in}}=\frac{\hat{X}^{\textsf{in}}+i\hat{P}^{\textsf{in}}}{\sqrt{2}}$, $\hat{C}^{\textsf{in}}_{j}=\frac{\hat{C}^{\textsf{in}}_{j,X}+i\hat{C}^{\textsf{in}}_{j,P}}{\sqrt{2}}$, and $\hat{D}^{\textsf{in}}_{j}=\frac{\hat{D}^{\textsf{in}}_{j,X}+i\hat{D}^{\textsf{in}}_{j,P}}{\sqrt{2}}$.

\section{SNR per photon}
We now introduce the figure of merit that evaluates the performance of sensing.

From the input-output theory, the output field $\hat{B}^{\textsf{out}}(t) $  reads 
\begin{flalign}
\begin{aligned}
\hat{B}^{\textsf{out}}(t)=\beta+\hat{B}^{\textsf{in}}(t)+\sqrt{\kappa}\hat{a}_1(t).
\end{aligned}
\end{flalign}
To estimate the perturbation $\epsilon$, we should integrate the output field over a long time period $[0, \tau]$.~The corresponding temporal mode is defined by \begin{flalign}
\begin{aligned}
\hat{\mathcal{B}}=\frac{1}{\sqrt{\tau}}\int^\tau_0\hat{B}^{\textsf{out}}(t)~dt,
\end{aligned}
\end{flalign}
which is a canonical bosonic annihilation operator.
For the perturbation $\epsilon\hat{V}=\epsilon\hat{a}^\dag_N\hat{a}_N$,  if  the drive $|\beta|\gg 1$, then the optimal observable is
\begin{flalign}
\begin{aligned}
\hat{\mathcal{M}}=\frac{1}{\sqrt{2}i}\Big{(}\hat{\mathcal{B}}-\hat{\mathcal{B}}^\dagger\Big{)},
\end{aligned}
\end{flalign}
which is  the $\hat{p}$-quadrature of the temporal output field $\hat{\mathcal{B}}$ \cite{McDonald2020}.

Let us first consider the case when $\epsilon$ is infinitesimal. Define the signal power in terms of the optimal observable $\hat{\mathcal{M}}$ as
\begin{flalign}\label{signalpower}
\begin{aligned}
\mathcal{S}(\epsilon)=|\langle\hat{\mathcal{M}}\rangle_\epsilon-\langle\hat{\mathcal{M}}\rangle_0|^2,
\end{aligned}
\end{flalign}
and the noise power as 
\begin{flalign}
\begin{aligned}
\mathcal{N}(\epsilon)=\langle\hat{\mathcal{M}}^2\rangle_{\epsilon}-\langle\hat{\mathcal{M}}\rangle_{\epsilon}^2. \end{aligned}
\end{flalign}
Here, the average $\langle\cdot\rangle_\epsilon$ represents the mean with  the steady state whose dynamics is governed by $\hat{H}[\epsilon]$. Since $\epsilon$ is infinitesimal, we can only consider the zeroth order of $\epsilon$ for the noise power. The  SNR is defined by 
\begin{flalign}
\begin{aligned}
\textrm{SNR}(\epsilon)=\frac{\mathcal{S}(\epsilon)}{\mathcal{N}(0)}.
\end{aligned}
\end{flalign}
Since the dominant term of the $\textrm{SNR}(\epsilon)$ with respect to $\epsilon$ is the same as that of the quantum Fisher information when $|\beta|\gg1$ \cite{Bao2021,McDonald2020,Baoli2021,Lau2018}, below we use the $\textrm{SNR}(\epsilon)$ to evaluate the performance of NH sensors.

To make a fair comparison,  the resources used in the measurement should be constrained. Following \cite{Baoli2021,McDonald2020} we take the SNR per photon denoted by 
\begin{flalign}
\begin{aligned}
\overline{\textrm{SNR}}(\epsilon)=\frac{\textrm{SNR}(\epsilon)}{\bar{n}_{\textsf{tot}}(0)}
\end{aligned}
\end{flalign}
as the figure of merit, where the total average photon number is 
\begin{flalign}
\begin{aligned}
\bar{n}_{\textsf{tot}}(0)= \sum_{n} \langle\hat{a}_n^\dagger\hat{a}_n\rangle_0\simeq \sum_n \langle\hat{a}_n^\dagger\rangle_0\langle\hat{a}_n\rangle_0
\end{aligned}
\end{flalign}
in the large-drive limit. Following the same reasoning as that of the noise power, only the zeroth order of $\bar{n}_{\textsf{tot} }$  in $\epsilon$ is concerned.

\section{$\overline{\textrm{SNR}}$ of NH sensors}

We now derive the $\overline{\textrm{SNR}}$ for NH sensors.
In the following, we take the perturbation Hamiltonian $\hat{V}$ in Eq.~(\ref{main}) as $\hat{V}=\hat{a}_N^\dagger\hat{a}_N$, and let the number of modes $N$ be odd. When $N$ is even, the scaling of $\overline{\textrm{SNR}}$ in terms of $A$ and $N$ is the same as that when $N$ is odd, except that the corresponding preceding multiplicative factors are different. 

We derive  the signal power $\mathcal{S}$, noise power $\mathcal{N}$, and the total average photon number $\bar{n}_{\textsf{tot}}$ in the presence of loss and gain as follows (see Appendix~\ref{appendixB} for details):
\begin{equation}\label{SNn}
\begin{aligned}\begin{aligned}
\mathcal{S}(\epsilon)
=&2\epsilon^2\kappa^2\beta^2\tau\cdot|{Q_{N,1}^\mathbb{X}}|^2\cdot|{Q_{1,N}^\mathbb{P}}|^2,\\
\mathcal{N}(0)=&\frac{1}{2}(1+\kappa Q_{1,1}^\mathbb{P})^2+\kappa\big{[} Q^\mathbb{P}(YY^\top+ZZ^\top){Q^\mathbb{P}}^{\top}\big{]}_{1,1} ,\\
\bar{n}_{\textsf{tot}}(0)=&\kappa\beta^2\big{[}{Q^\mathbb{X}}^{\top}Q^\mathbb{X}\big{]}_{1,1},
\end{aligned} \end{aligned}
\end{equation}
with information matrices 
\begin{flalign}
\begin{aligned}
Q^{\mathbb{X}}=(h^{\mathbb{X}}+YY^\top-ZZ^\top)^{-1}
\end{aligned}
\end{flalign}
and 
\begin{flalign}
\begin{aligned}
Q^{\mathbb{P}}=(h^{\mathbb{P}}+YY^\top-ZZ^\top)^{-1}.
\end{aligned}
\end{flalign}

In the absence of  loss and gain, namely  $Z=0$ and $Y=0$,  it was demonstrated in \cite{McDonald2020} that $\overline{\textrm{SNR}}(\epsilon)\propto \exp\{2A(N-1)\}$ implying that an  exponentially enhanced sensitivity can be obtained. The key idea is illustrated in Fig.~1(b).  To detect  $\epsilon$,  a real drive is injected at  site 1 to excite the \textrm{X} chain, then the wavepacket propagates  rightwards. When it  reaches the last site, the signal power grows with a factor of $e^{2A(N-1)}$. Then at site $N$, due to the perturbation, the wavepacket scatters off the boundary and changes to  $\hat{p}_N$ quadrature. It then propagates backwards to site 1 amplifying the signal. If the $\hat{p}$-quadrature  of the output field is measured, then a total amplification factor $e^{4A(N-1)}$ of the signal power is obtained. While for the total average photon number, it amplifies only along one traversal of the chain obtaining an amplification factor of $e^{2A(N-1)}$. As for the noise power, for the ideal case of zero internal loss and gain, the noise power is the same as that of the input field, namely, $\mathcal{N}(0)=1/2$. Combining this with the amplification factors of the signal power and the total average photon number, the exponentially large factor  $e^{2A(N-1)}$ of $\overline{\textrm{SNR}}(\epsilon)$ can be explained. 

In the presence of loss and gain, owing to the nonreciprocal dynamics governed by  $(Q^\mathbb{X})^{-1}$ and $(Q^\mathbb{P})^{-1}$, the noise power may be significantly amplified in general, which satisfies $\mathcal{N}\propto e^{2A(N-1)}$, causing the vanishing of the ideally exponential sensitivity. In addition, from Eq.~(\ref{main}), the net noise matrix ($YY^\top-ZZ^\top$)  may cause the sensing dynamics to become unstable. This  implies that noise may lead to at least one of the eigenvalues of the noisy NH dynamical matrix  $(h^\mathbb{X}+YY^\top-ZZ^\top)$ or $(h^\mathbb{P}+YY^\top-ZZ^\top)$ to sit in the right half plane. In this case, Eq.~(\ref{main}) has no steady state and the expectation of some of the canonical quadratures will diverge to infinity, which is physically meaningless \cite{Franklin2019} (see Appendix~\ref{appendixC} for details).

\section{Tuning loss and gain}
We now present how to achieve exponentially enhanced and stable NH sensing by proactively tuning the loss and gain.

It is widely believed that introducing gain is necessary to address the sensitivity revival problem in the presence of loss. However, we find that the loss  $Z$ plays a pivotal role.  To  this end, consider the case where there is only loss and no gain, namely, $Y=0$. Given the dynamical matrix $h^{\mathbb{P}}$, we prove that if the loss couplings $Z$ can be tuned such that all its columns lie in the linear space spanned by the second column $h^{\mathbb{P}}_{\bm\cdot 2}$ through the last column $h^{\mathbb{P}}_{\bm\cdot N}$ of the dynamical matrix $h^{\mathbb{P}}$, then we can revive the ideally exponential sensitivity when the sensing dynamics is stable (see Appendix~\ref{appendixD} for details).  In short, to attain an exponential sensitivity,  the loss coupling matrix  $Z$ should meet
\begin{description}
\item[$\text{(\textbf{C1})}$] $\textbf{col}(Z)\subseteq \textbf{Span}\big{\{}h^{\mathbb{P}}_{\bm\cdot 2}, ~h^{\mathbb{P}}_{\bm\cdot 3},  \cdots, h^{\mathbb{P}}_{\bm\cdot N}\big{\}}.$
\end{description}
This finding is remarkable as it defies the general belief that even a weak loss will quickly lead to vanishing of quantum advantages in high precision sensing.

\begin{figure}[h!]
\centering
\includegraphics[scale=0.62]{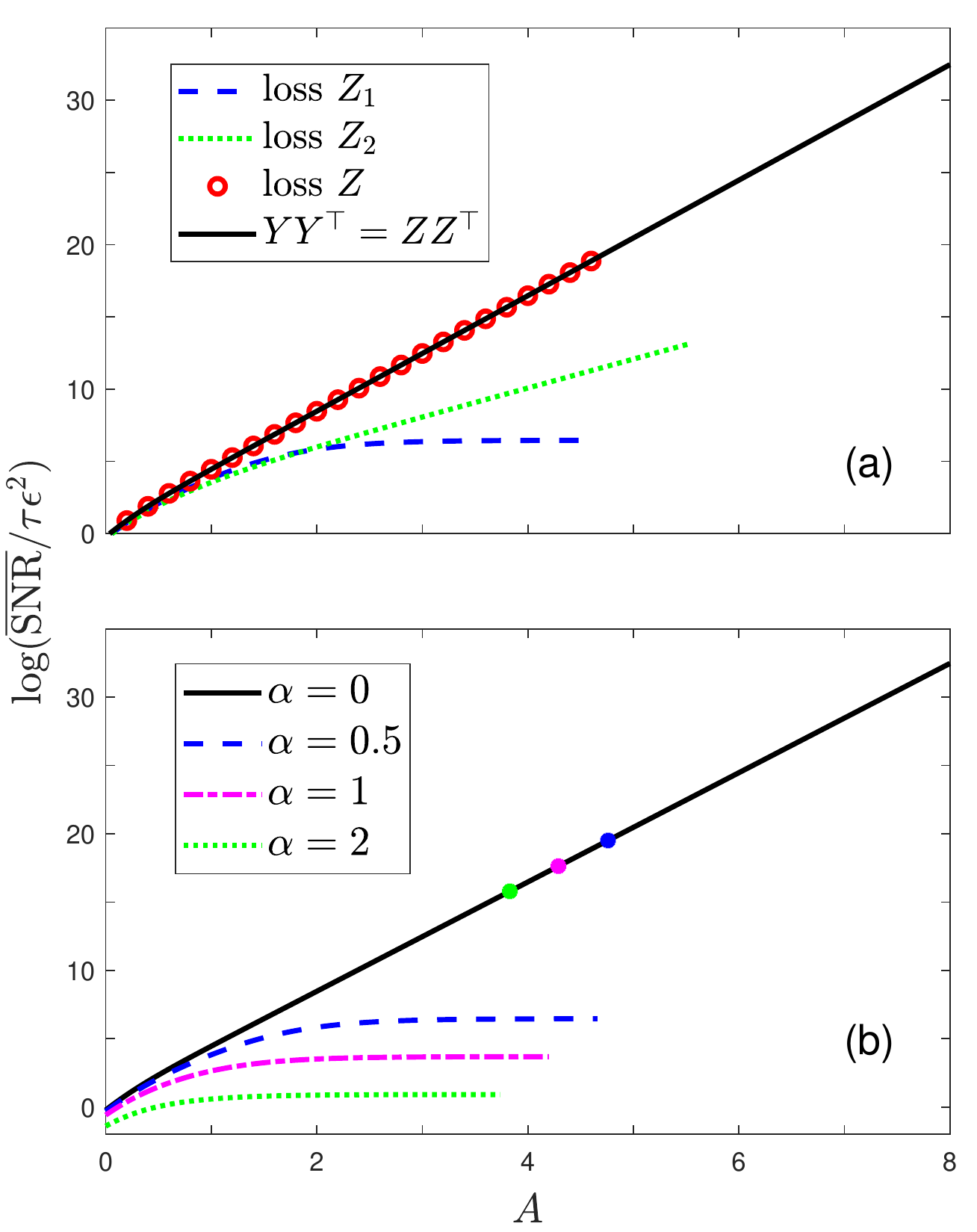}
\caption{(Color online) The performance $\log(\frac{\overline{\textrm{SNR}}}{\tau\epsilon^2})$ versus $A$ under different loss couplings and $\alpha$'s. The range of $A$ is identified to guarantee a stable dynamics.  (a) $\alpha=0.5$. Dashed blue: loss  $Z_1$, dotted green: loss $Z_2$, red circle: loss $Z$  which satisfies  ($\mathbf{C1}$), black line: the ideal case and the tuned case where both  ($\mathbf{C1}$) and ($\mathbf{C2}$) are met. (b) For loss $Z_1$, black solid: $\alpha=0$, blue dashed: $\alpha=0.5$, magenta dot-dashed: $\alpha=1$, green dotted: $\alpha=2$. The best sensitivity that can be revived under loss $Z$ for different values of $\alpha$ is shown by the colored dots on the ideal black solid line.}
\end{figure}

To illustrate the above, consider a simplest 3-site NH sensor. In the ideal case, namely there is no loss and gain, the ideal $\overline{\textrm{SNR}}\propto e^{4A}$ as depicted by the black line in Fig.~2(a). Assume now that there are two loss baths in total ($N_Z=2$) and consider two different loss couplings described by  
\begin{flalign}
\begin{aligned}
Z_1=\alpha\left(
 \begin{array}{ccc}
 -e^A & -e^A  \\
 0 & 1  \\
 e^{-A} & 0  \\
 \end{array}
 \right)
\end{aligned} 
\end{flalign}
 and 
\begin{flalign}
\begin{aligned}
Z_2=\alpha\left(
 \begin{array}{ccc}
 -e^A & 0  \\
 0 & 1  \\
 e^{-A} & e^{-A} \\
 \end{array}
 \right),
\end{aligned} 
\end{flalign}
respectively. It is clear that $Z_1$ and $Z_2$ do not obey  \text{(\textbf{C1})}. With parameters $\alpha=0.5$, $\kappa=10$, $\omega=10^5$ and $J=\omega\frac{2e^A}{e^{2A}+1}$,   we depict the
 $\overline{\textrm{SNR}}$ under $Z_1$ (dashed blue) and $Z_2$ (dotted green) in Fig.~2(a). Here, we only care about the amplification factor $A$'s that make the dynamics stable. Note that the  $\overline{\textrm{SNR}}$ under $Z_1$  approaches a constant as $A$ increases, while under $Z_2$ the $\overline{\textrm{SNR}}$ grows as $A$ increases, but is still much smaller than the ideal case.  In Fig.~2(b), we illustrate the $\overline{\textrm{SNR}}$ under $Z_1$ with different $\alpha$'s.  If $\alpha=0$, it is the ideal case (black). 
 
To revive the ideally exponential sensitivity, for $Z_1$ we  proactively add an {\it exponentially small}  coupling  $Z_{32}=e^{-A}$  between the third mode and the second loss bath, while for $Z_2$ we  proactively add an {\it exponentially large} coupling $Z_{12}=-e^{A}$ between the first mode and the second bath.  The resulting loss coupling matrix becomes \begin{flalign} 
\begin{aligned}
Z=\alpha\left(
 \begin{array}{ccc}
 -e^A & -e^A  \\
 0 & 1 \\
 e^{-A} & e^{-A}\\
 \end{array}
 \right)
 \end{aligned} 
\end{flalign}
which satisfies (\textbf{C1}).  The
 $\overline{\textrm{SNR}}$ under $Z$ (red circle) is illustrated in Fig.~2(a). It is clear that the ideally exponential sensitivity is regained in the stable region and the best $\overline{\textrm{SNR}}$ obtained under $Z$ is much better than those attained under $Z_1$ and $Z_2$. 
When tuning $Z_1$ to be $Z$, for different values of $\alpha$, the range of $A$ that makes the system stable
is different, so is the best sensitivity that can be regained. In Fig.~2(b), we depict the corresponding best sensitivity that can be attained by colored dots on the ideal black line for different $\alpha$'s.  It is clear that, as $\alpha$ increases, the best sensitivity that can be attained decreases.

To address the instability problem and {\it fully} regain the ideally exponential sensitivity, we can further introduce gain, and tune $Y$ such that it meets the balanced condition
\begin{description}
\item[$\text{(\textbf{C2})}$] $YY^\top=ZZ^\top.$
\end{description}
We prove that under (\textbf{C1}) and (\textbf{C2}), an exponential enhancement  can be fully revived for noisy NH sensing, that is, $\overline{\textrm{SNR}} \propto e^{2A(N-1)}$  (see Appendix~\ref{appendixE} for details). In this case, the range of $A$ in Fig.~2 can, in principle, be arbitrarily large, which of course depends on the parameters of the real setup.

We now consider the robustness of the sensitivity under  (\textbf{C1}) and (\textbf{C2}).   To this end, we can first reverse the roles played by the loss coupling matrices $Z$ and $Z_1$. Assume that when tuning the desired loss $Z$,  we obtain $Z_1$ instead of $Z$, namely the desired loss rate  $Z_{32}=e^{-A}$ is set to be $0$ in practice. From Fig.~2(a), it is clear that even though there is only an exponentially small imperfection, the sensitivity can be greatly reduced when $A\in[2,4.6]$. 

Second, assuming  
\begin{flalign}
(YY^\top-ZZ^\top)=\gamma(|1\rangle\langle N|+| N\rangle\langle1|), 
\end{flalign}
it can be verified that to ensure a stable  sensing dynamics,  a necessary condition for $\gamma$ is   $|\gamma|< \kappa e^{-A(N-1)}$ (see Appendix~\ref{appendixC} for details). In this case, there is a striking tradeoff between the enhancement of the sensitivity and the exponential decrement of the robust stability. This owes to the fact that there is ``no free lunch". For a sensor having exponential sensitivity, it must have an inherent highly nonlinear amplification mechanism, such as that described by $h^{\mathbb{P}}$. Thus, fine tunings at certain key points are inevitable. Otherwise, the residual noise may either be greatly amplified or lead to system instability, both of which will severely degrade the sensitivity. Using a similar analysis to the above, we can determine which coupling rates of loss and gain need to be finely tuned. In this way, we can adjust these couplings as well as possible to construct a noisy NH sensor with ultra-high sensitivity.

The feasibility of our approach is well supported by the current capability of engineering loss and gain in optics in a controlled manner \cite{Feng2014,Liuyl2017,Ren2022}.
By proactively tuning the loss and gain couplings, our proposal opens a new way to significantly enhance the sensitivity in the presence of loss and gain.

\section{Regime beyond linear response}
We now relax the assumption that the perturbation is infinitesimal. When the parameter to be detected, $\epsilon_0$, is not infinitely small, not only the linear response of $\epsilon_0$, but all orders in $\epsilon_0$ of the output field should be calculated.

As in  \cite{Baoli2021,McDonald2020}, we focus on the most interesting case where $\epsilon_0/\kappa\ll1$. Take
\begin{flalign}
\begin{aligned} 
\overline{\textrm{SNR}}(\epsilon_0)=\frac{\mathcal{S}(\epsilon_0)}{(\mathcal{N}\cdot\bar{n}_{\textsf{tot}})}
\end{aligned} 
\end{flalign}
as the figure of merit, which quantifies the distinguishability between Gaussian homodyne current distributions under $\epsilon=0$ and $\epsilon=\epsilon_0$. Here, the noise power
$\mathcal{N}=[\mathcal{N}(0)+\mathcal{N}(\epsilon_0)]/2,$
and the total average number of photons
$\bar{n}_{\textsf{tot}}=[\bar{n}_{\textsf{tot}}(0)+\bar{n}_{\textsf{tot}}(\epsilon_0)]/2.$

With the perturbation $\epsilon_0\hat{V}=\epsilon_0\hat{a}_N^\dagger\hat{a}_N$, we verify that  under the balanced  condition (\textbf{C2}) and the following two conditions,
\begin{description}
\item[$\text{(\textbf{C3})}$] $\textbf{col}(Z)\subseteq \textbf{Span}\big{\{}h^{\mathbb{P}}_{\bm\cdot 2},~ h^{\mathbb{P}}_{\bm\cdot 3},  \cdots, h^{\mathbb{P}}_{\bm\cdot (N-1)}\big{\}},$
\item[$\text{(\textbf{C4})}$] $\textbf{col}(Z) \perp (h^\mathbb{X})^{-1}_{N \bm\cdot}$,
\end{description}
the best revived sensitivity is the same as that in \cite{McDonald2020} where there is no loss and gain (see Appendix~\ref{appendixF} for details). 

Note that in the regime beyond linear response, to revive the ideal sensitivity, \text{(\textbf{C3})} and \text{(\textbf{C4})} are slightly stricter than \text{(\textbf{C1})}, which constrains the loss  in  linear response. Here, \text{(\textbf{C3})} means that the columns of $Z$ should reside in the linear space generated by the second column through the $(N-1)$th column of the dynamical matrix $h^{\mathbb{P}}$, while  \text{(\textbf{C4})} implies that the columns of $Z$ should be  orthogonal to the $N$th row of the ideal information matrix  $(h^\mathbb{X})^{-1}$.

% \begin{equation*}
%\begin{aligned}
%&\mathcal{S}(\epsilon_0)= 2\tau\kappa^2\beta^2\bigg(\mathbb{H}[\epsilon_0]^{-1}_{N+1,1}-\mathbb{H}[0]^{-1}_{N+1,1}\bigg)^2,\\
%%=&2\tau\kappa^2\beta^2\frac{\epsilon_0^2}{(\frac{\kappa^2}{4}+\epsilon_0^2)^2}e^{4A(N-1)},\\
%&\mathcal{N}=\frac{1}{4}\Bigg\{(1+\kappa(h^{\mathbb{P}})^{-1}_{1,1})^2+(1+\kappa\mathbb{H}[\epsilon_0]^{-1}_{N+1,N+1})^2\\
%&~~~~~\kappa^2(\mathbb{H}[\epsilon_0]^{-1}_{\!N+1\!,\!1})^2\!+\!4\kappa\big[(h^{\mathbb{P}})^{-1}YY^\top (h^{\mathbb{P}})^{-1\top}\big]_{1,1}\\
%&~~~~+\!4\kappa\!\bigg[\mathbb{H}[\epsilon_0]^{-\!1}\!\!\begin{pmatrix}
%                                                                                                                                                                     YY^\top & \! 0 \\
%                                                                                                                                                                    \! 0 & \! YY^\top
%                                                                                                                                                                   \end{pmatrix}\!\!(\mathbb{H}[\epsilon_0]^{-\!1}\!)^\top\!\bigg]_{\!N\!+\!1,\!N\!+\!1}\Bigg\},\\
%&\bar{n}_{\textsf{tot}}=\frac{1}{2}\kappa\beta^2\sum_{n=1}^{N}\Big(\mathbb{H}[0]^{-1}_{n,1})^2+(\mathbb{H}[\epsilon_0]^{-1}_{n,1})^2\\
%&~~~~~~~~~~~~~~~~~~~~~~~~~+(\mathbb{H}[\epsilon_0]^{-1}_{N+n,1}\Big)^2.
%\end{aligned}
%\end{equation*}
%}
\section{Conclusion}

We have investigated the ideal sensitivity revival and the  stability of noisy NH quantum sensing. We present a strategy to proactively tune the loss and gain couplings to construct a stable NH sensor achieving an exponential sensitivity. We find that the loss is key to revive the sensitivity, and that  balanced gain and loss are crucial to fully regain the ideal sensitivity and to ensure a stable NH sensor, no matter if the parameter is infinitesimal or in the regime beyond linear response. We also point out that to design a  noisy sensor with ultra-high sensitivity, fine tunings are inevitable at certain key points. Our proposal opens a new way to enhance the sensitivity of  noisy sensors by proactively tuning the loss and gain, and may have potential applications in quantum sensing and quantum engineering.

\section*{Acknowledgments}
%L.B. acknowledges the support of the Fundamental Research Funds for the Central Universities of China (No. 3122023QD23). B.Q. acknowledges the support of the National Natural Science Foundation of China (No.~61773370). F.N. is supported in
%part by: Nippon Telegraph and Telephone Corporation (NTT) Research, the Japan Science and Technology Agency (JST) [via the Quantum Leap Flagship Program (Q-LEAP), and the Moonshot R$\&$D Grant Number JPMJMS2061],
%the Asian Office of Aerospace Research and Development (AOARD) (via Grant No. FA2386-20-1-4069),
%and the Office of Naval Research (ONR) Global (via Grant No. N62909-23-1-2074). D.D. acknowledges the support of the Australian Research Council Future Fellowship funding scheme under Project FT220100656.

L.B.~acknowledges the support of the Fundamental Research Funds for the Central Universities of China (No.~3122023QD23).
B.Q. acknowledges the support of the National Natural Science Foundation of China (No.~61773370). D.D. acknowledges the support of  the Australian Research Council Future Fellowship funding scheme under Project FT220100656. F.N. is supported in part by:
Nippon Telegraph and Telephone Corporation (NTT) Research,
the Japan Science and Technology Agency (JST)
[via the Quantum Leap Flagship Program (Q-LEAP), and the Moonshot R$\&$D Grant Number JPMJMS2061],
the Asian Office of Aerospace Research and Development (AOARD) (via Grant No. FA2386-20-1-4069),
and the Office of Naval Research (ONR) Global (via Grant No. N62909-23-1-2074).

\clearpage

\begin{widetext}

\section*{Appendices}
To make the paper self-contained, this appendices are organized as follows. We first describe the total Hamiltonian of the sensor and derive the Heisenberg-Langevin equations in Appendix~\ref{appendixA} . Then we derive the signal-to-noise ratio (SNR) per photon in Appendix~\ref{appendixB}. The real matrix $Z~(Y)$ describes the coupling between the system and the loss (gain) bath.  In Appendix~\ref{appendixC} we derive the necessary conditions to ensure the stability of the dynamics if the loss and gain are unbalanced. In Appendix~\ref{appendixD} we prove that when the gain coupling $Y=0$, the loss coupling $Z$ satisfies condition $\mathbf{(C1)}$, and if the dynamics is stable, the signal power $\mathcal{S}$, noise power $\mathcal{N}$, and the total average photon number $\bar{n}_{\textsf{tot}}(0)$ are the same as those of the ideal noise-free case. The SNR per photon under conditions $(\mathbf{C1})$ and $(\mathbf{C2})$ are given in Appendix~\ref{appendixE}. In Appendix~\ref{appendixF} we calculate the SNR per photon in the regime beyond linear response. The calculations of  the elements of $\mathbb{H}[\epsilon]^{-1}$ and $\mathbb{H}[\epsilon_0]^{-1}$ are shown in Appendix~\ref{appendixG} and Appendix~\ref{appendixH}, respectively.

\appendix

\section{The non-Hermitian sensor and the Heisenberg-Langevin equations}\label{appendixA}
The total Hamiltonian of the sensor is described by
\begin{equation}
\begin{aligned}
\hat{H}_{\text{tot}}=&\hat{H}_S+\hat{H}_{\epsilon}+\hat{H}_{\textrm{input}}+\hat{H}_{\textrm{wave}}+\hat{H}_{\textrm{gain}}+\hat{H}_{\textrm{loss}}+\hat{H}_{S,\textrm{wave}}+\hat{H}_{S,\textrm{gain}}+\hat{H}_{S,\textrm{loss}},
\end{aligned}
\end{equation}
with the perturbation Hamiltonian
\begin{equation}
\begin{aligned}
\hat{H}_{\epsilon}=\epsilon\hat{V},
\end{aligned}
\end{equation}
input Hamiltonian
\begin{equation}
\begin{aligned}
\hat{H}_{\textrm{input}}=-i\sqrt{\kappa}~(\hat{a}_1^\dagger \beta-\hat{a}_1 \beta^\dagger),
\end{aligned}
\end{equation}
waveguide Hamiltonian
\begin{equation}
\begin{aligned}
\hat{H}_{\textrm{wave}}=\int d k ~(k \hat{b}_k^\dagger \hat{b}_k),
\end{aligned}
\end{equation}
the $j$th gain bath Hamiltonian
\begin{equation}
\begin{aligned}
\hat{H}_{\textrm{gain}}=\int d k ~(k \hat{c}_{j,k}^\dagger \hat{c}_{j,k}),
\end{aligned}
\end{equation}
the $j$th loss bath Hamiltonian
\begin{equation}
\begin{aligned}
\hat{H}_{\textrm{loss}}=\int d k ~(k \hat{d}_{j,k}^\dagger \hat{d}_{j,k}),
\end{aligned}
\end{equation}
the interaction Hamiltonian between the chain and the waveguide
\begin{equation}
\begin{aligned}
\hat{H}_{S,\textrm{wave}}=\int dk~ \frac{1}{\sqrt{\pi}}\sqrt{\frac{\kappa}{2}}~(\hat{a}_1 \hat{b}_k^\dagger +\hat{a}_1^\dagger \hat{b}_k),
\end{aligned}
\end{equation}
the interaction Hamiltonian between the chain and the $j$th gain bath
\begin{equation}
\begin{aligned}
\hat{H}_{S,\textrm{gain}}=\sum_{i=1}^N\sum_{j=1}^{N_Y}\int dk~ \frac{1}{\sqrt{\pi}}Y_{i,j}(\hat{a}_i\hat{c}_{j,k}+\hat{a}_i^\dagger\hat{c}_{j,k}^\dagger),
\end{aligned}
\end{equation}
and
the interaction Hamiltonian between the chain and the $j$th loss bath
\begin{equation}
\begin{aligned}
\hat{H}_{S,\textrm{loss}}=\sum_{i=1}^N\sum_{j=1}^{N_Z}\int dk~ \frac{1}{\sqrt{\pi}}Z_{i,j}(\hat{a}_i\hat{d}^\dagger_{j,k}+\hat{a}_i^\dagger\hat{d}_{j,k}).
\end{aligned}
\end{equation}
Here, $\hat{a}_i$ denotes the mode annihilation operator on site $i$, $\hat{b}_k$ is the annihilation operator of the mode with wave number $k$ in the waveguide, $\hat{c}_{j,k}$ is the annihilation operator of the $j$th gain bath with wave number $k$, and $\hat{d}_{j,k}$ is the annihilation operator of the $j$th loss bath mode with wave number $k$. The real matrix $Z$ ($Y$) depicts the coupling between the system and the loss (gain) bath.

The Heisenberg equations of  motion for the cavity modes and the field modes are
\begin{equation}\label{abcdmotion}
\begin{aligned}
\frac{d\hat{a}_n}{dt}=&~w \hat{a}_{n-1}+\Delta\hat{a}^\dagger_{n+1}+\Delta\hat{a}^\dagger_{n-1}-w\hat{a}_{n+1}-i\epsilon[\hat{a}_n,\hat{V}]-\sqrt{\kappa}\beta\delta_{n,1}\\
&-i\delta_{n,1}\int dk~ (\frac{1}{\sqrt{\pi}}\sqrt{\frac{\kappa}{2}} \hat{b}_k)-i\sum_{j=1}^{N_Y}\int dk~(\frac{1}{\sqrt{\pi}}Y_{n,j}\hat{c}_{j,k}^\dagger)-i\sum_{j=1}^{N_Z}\int dk~(\frac{1}{\sqrt{\pi}}Z_{n,j}\hat{d}_{j,k}),\\
\frac{d\hat{b}_k}{dt}=&-ik\hat{b}_k-i\frac{1}{\sqrt{\pi}}\sqrt{\frac{\kappa}{2}} \hat{a}_1,\\
\frac{d\hat{c}_{j,k}}{dt}=&-i k\hat{c}_{j,k}-i\sum_{i=1}^N\frac{1}{\sqrt{\pi}}Y_{i,j}\hat{a}_i^\dagger,\\
\frac{d\hat{d}_{j,k}}{dt}=&-i k\hat{d}_{j,k}-i\sum_{i=1}^N\frac{1}{\sqrt{\pi}}Z_{i,j}\hat{a}_i.
\end{aligned}
\end{equation}
The solutions of the last three equations in Eq. \eqref{abcdmotion} are
\begin{equation}\label{bcdmotion}
\begin{aligned}
\hat{b}_k=&~e^{-ik(t-t_0)}\hat{b}_k(t_0)-i\frac{1}{\sqrt{\pi}}\sqrt{\frac{\kappa}{2}}\int_{t_0}^t dt' e^{-i k (t-t')}\hat{a}_1(t'),\\
\hat{c}_{j,k}=&~e^{-ik(t-t_0)}\hat{c}_{j,k}(t_0)-i\frac{1}{\sqrt{\pi}}\sum_{i=1}^N\int_{t_0}^t dt' e^{-i k (t-t')}Y_{ij}\hat{a}_i^\dagger(t'),\\
\hat{d}_{j,k}=&~e^{-ik(t-t_0)}\hat{d}_{j,k}(t_0)-i\frac{1}{\sqrt{\pi}}\sum_{i=1}^N\int_{t_0}^t dt' e^{-i k (t-t')}Z_{ij}\hat{a}_i(t').
\end{aligned}
\end{equation}
Substituting Eq. \eqref{bcdmotion} into the first equation of  Eq.~\eqref{abcdmotion} yields
\begin{equation}
\begin{aligned}
\frac{d\hat{a}_n}{dt}=&~w \hat{a}_{n-1}+\Delta\hat{a}^\dagger_{n+1}+\Delta\hat{a}^\dagger_{n-1}-w\hat{a}_{n+1}-i\epsilon[\hat{a}_n,\hat{V}]-\sqrt{\kappa}\beta\delta_{n,1}\\
&-i\sqrt{\frac{\kappa}{2\pi}}\delta_{n,1}\int dk e^{-ik(t-t_0)}\hat{b}_k(t_0)-\delta_{n,1}\frac{\kappa}{2\pi}\int dk \int_{t_0}^t dt' e^{-i k (t-t')}\hat{a}_1\\
&-i\sum_{j=1}^{N_Y}\int dk \frac{1}{\sqrt{\pi}}Y_{n,j}e^{ik(t-t_0)}\hat{c}_{j,k}^\dagger(t_0)+\frac{1}{\pi}\sum_{j=1}^{N_Y}\sum_{i=1}^NY_{n,j}Y_{i,j}\int dk\int ^t_{t_0}dt'e^{ik(t-t')}\hat{a}_i\\
&-i\sum_{j=1}^{N_Z}\int dk \frac{1}{\sqrt{\pi}}Z_{n,j}e^{-ik(t-t_0)}\hat{d}_{j,k}(t_0)+\frac{1}{\pi}\sum_{j=1}^{N_Z}\sum_{i=1}^NZ_{n,j}Z_{i,j}\int dk\int ^t_{t_0}dt'e^{-ik(t-t')}\hat{a}_i\\
=&~w \hat{a}_{n-1}+\Delta\hat{a}^\dagger_{n+1}+\Delta\hat{a}^\dagger_{n-1}-w\hat{a}_{n+1}-i\epsilon[\hat{a}_n,\hat{V}]
-\sqrt{\kappa}\beta\delta_{n,1}\\
&-\frac{\kappa}{2}\hat{a}_1(t)\delta_{n,1}+\sum_{i=1}^N(YY^\top-ZZ^\top)_{n,i}\hat{a}_i-\sqrt{\kappa}\hat{B}^{\textsf{in}}\delta_{n,1}-\sqrt{2}~\Bigg(\sum_{j=1}^{N_{Y}}Y_{n,j}\hat{C}_{j}^{\textsf{in}\dagger}+\sum_{j=1}^{N_{Z}}Z_{n,j}\hat{D}_{j}^{\textsf{in}}\Bigg),
\end{aligned}
\end{equation}
where we have defined 
\begin{equation}
\begin{aligned}
\hat{B}^{\textsf{in}}&=i\sqrt{\frac{1}{2\pi}}\int dk e^{-ik(t-t_0)}\hat{b}_k(t_0), \\ \hat{C}^{\textsf{in}}_j&=i\sqrt{\frac{1}{2\pi}}\int dk e^{-ik(t-t_0)}\hat{c}_{j,k}(t_0),\\ \hat{D}^{\textsf{in}}&=i\sqrt{\frac{1}{2\pi}}\int dk e^{-ik(t-t_0)}\hat{d}_{j,k}(t_0),
\end{aligned}
\end{equation}
and used the equations $\int dk e^{-ik(t-t')}=2\pi \delta(t-t')$ and $\int_{t_0}^t dt' \delta(t-t')\hat{a}_i(t')=\frac{1}{2}\hat{a}_i(t)$. To ensure the Markovian nature of the entire dynamics, $\hat{B}^{\textsf{in}}$, $\hat{C}_{j}^{\textsf{in}}$ and $\hat{D}_{j}^{\textsf{in}}$ are assumed to be quantum Gaussian white noise: $\langle Q(t)Q^\dagger(t')\rangle=(\bar{n}^{\textsf{th}}_Q+1)\delta(t-t')$, $\langle Q^\dagger(t)Q(t')\rangle=\bar{n}^{\textsf{th}}_Q\delta(t-t')$, and $\langle Q(t)Q(t')\rangle=0$,
where $Q\in \{ \hat{B}^{\textsf{in}},~\hat{C}^{\textsf{in}}_j,~\hat{D}^{\textsf{in}}_j \}$, and there are no correlations between different noise operators. Here, $\bar{n}^{\textsf{th}}_Q$ is the number of thermal quanta in the input field. Therefore, the Heisenberg-Langevin equations can be expressed as
%Eq. \eqref{anheisenberg}.
\begin{equation}\label{anheisenberg}
\begin{aligned}
\frac{d\hat{a}_n}{dt}=&~w \hat{a}_{n-1}-w\hat{a}_{n+1}+\Delta\hat{a}^\dagger_{n+1}+\Delta\hat{a}^\dagger_{n-1}-i\epsilon[\hat{a}_n,\hat{V}]-\frac{\kappa}{2}\hat{a}_1\delta_{n,1}\\
&+\sum_{j=1}^{N_{Y}}\sum_{i=1}^{N}Y_{n,j}Y_{i,j}\hat{a}_i-\sum_{j=1}^{N_{Z}}\sum_{i=1}^{N}Z_{n,j}Z_{i,j}\hat{a}_i-\sqrt{\kappa}(\hat{B}^{\textsf{in}}+\beta)\delta_{n,1}-\sqrt{2}\;\Bigg(\sum_{j=1}^{N_{Y}}Y_{n,j}\hat{C}_{j}^{\textsf{in}\dagger}+\sum_{j=1}^{N_{Z}}Z_{n,j}\hat{D}_{j}^{\textsf{in}}\Bigg).
\end{aligned}
\end{equation}

To see  how the signal is amplified, it is better to turn to the picture of canonical quadratures $\hat{x}_n$ and $\hat{p}_n$ defined  via $\hat{a}_n=\frac{\hat{x}_n+i\hat{p}_n}{\sqrt{2}}$.
Then the corresponding  Heisenberg-Langevin equations in terms of $\hat{x}_n$ and $\hat{p}_n$ read
\begin{equation}
\begin{aligned}
\frac{d\hat{x}_n}{dt}=&-(w-\Delta)\hat{x}_{n+1}+(w+\Delta)\hat{x}_{n-1}-i\epsilon[\hat{x}_n,\hat{V}]-\frac{\kappa}{2}\hat{x}_1\delta_{n,1}\\
&+\sum_{j=1}^{N_{Y}}\sum_{i=1}^{N}Y_{n,j}Y_{i,j}\hat{x}_i-\sum_{j=1}^{N_{Z}}\sum_{i=1}^{N}Z_{n,j}Z_{i,j}\hat{x}_i\\
&-\sqrt{\kappa}\frac{\hat{B}^{\textsf{in}}+\hat{B}^{\textsf{in}\dagger}}{\sqrt{2}}\delta_{n,1}-\sqrt{2\kappa}\beta\delta_{n,1}-\sqrt{2}~\Bigg(\sum_{j=1}^{N_{Y}}Y_{n,j}\frac{\hat{C}_{j}^{\textsf{in}\dagger}+\hat{C}_{j}^{\textsf{in}}}{\sqrt{2}}+\sum_{j=1}^{N_{Z}}Z_{n,j}\frac{\hat{D}_{j}^{\textsf{in}}+\hat{D}_{j}^{\textsf{in}\dagger}}{\sqrt{2}}\Bigg),\\
\frac{d\hat{p}_n}{dt}=&~(w-\Delta)\hat{p}_{n-1}-(w+\Delta)\hat{p}_{n+1}-i\epsilon[\hat{p}_n,\hat{V}]-\frac{\kappa}{2}\hat{p}_1\delta_{n,1}\\
&+\sum_{j=1}^{N_{Y}}\sum_{i=1}^{N}Y_{n,j}Y_{i,j}\hat{p}_i-\sum_{j=1}^{N_{Z}}\sum_{i=1}^{N}Z_{n,j}Z_{i,j}\hat{p}_i\\
&-\sqrt{\kappa}\frac{\hat{B}^{\textsf{in}}-\hat{B}^{\textsf{in}\dagger}}{\sqrt{2}i}\delta_{n,1}-\sqrt{2}~\Bigg(\sum_{j=1}^{N_{Y}}Y_{n,j}\frac{\hat{C}_{j}^{\textsf{in}\dagger}-\hat{C}_{j}^{\textsf{in}}}{\sqrt{2}i}+\sum_{j=1}^{N_{Z}}Z_{n,j}\frac{\hat{D}_{j}^{\textsf{in}}-\hat{D}_{j}^{\textsf{in}\dagger}}{\sqrt{2}i}\Bigg).
\end{aligned}
\end{equation}
By defining $\hat{B}^{\textsf{in}}=\frac{\hat{X}^{\textsf{in}}+i\hat{P}^{\textsf{in}}}{\sqrt{2}}$, $\hat{C}_{j}^{\textsf{in}}=\frac{\hat{C}^{\textsf{in}}_{j,X}+i\hat{C}^{\textsf{in}}_{j,P}}{\sqrt{2}}$, $\hat{D}_{j}^{\textsf{in}}=\frac{\hat{D}^{\textsf{in}}_{j,X}+i\hat{D}^{\textsf{in}}_{j,P}}{\sqrt{2}}$, and let $J= \sqrt{w^2-\Delta^2}$ and $\exp\{2A\}= \frac{w+\Delta}{w-\Delta}$, the above equation can be described by
\begin{equation}
\begin{aligned}
\frac{d\hat{x}_n}{dt}=&-Je^{-A}\hat{x}_{n+1}+Je^A\hat{x}_{n-1}-i\epsilon[\hat{x}_n,\hat{V}]-\frac{\kappa}{2}\hat{x}_1\delta_{n,1}\\
&+\sum_{j=1}^{N_{Y}}\sum_{i=1}^{N}Y_{n,j}Y_{i,j}\hat{x}_i-\sum_{j=1}^{N_{Z}}\sum_{i=1}^{N}Z_{n,j}Z_{i,j}\hat{x}_i\\
&-\sqrt{\kappa}\hat{X}^{\textsf{in}}\delta_{n,1}-\sqrt{2\kappa}\beta\delta_{n,1}-\sqrt{2}~\Big(\sum_{j=1}^{N_{Y}}Y_{n,j}\hat{C}^{\textsf{in}}_{j,X}+\sum_{j=1}^{N_{Z}}Z_{n,j}\hat{D}^{\textsf{in}}_{j,X}\Big),\\
\frac{d\hat{p}_n}{dt}=&~Je^{-A}\hat{p}_{n-1}-Je^A\hat{p}_{n+1}-i\epsilon[\hat{p}_n,\hat{V}]-\frac{\kappa}{2}\hat{p}_1\delta_{n,1}\\
&+\sum_{j=1}^{N_{Y}}\sum_{i=1}^{N}Y_{n,j}Y_{i,j}\hat{p}_i-\sum_{j=1}^{N_{Z}}\sum_{i=1}^{N}Z_{n,j}Z_{i,j}\hat{p}_i\\
&-\sqrt{\kappa}\hat{P}^{\textsf{in}}\delta_{n,1}-\sqrt{2}~\Big(-\sum_{j=1}^{N_{Y}}Y_{n,j}\hat{C}^{\textsf{in}}_{j,P}+\sum_{j=1}^{N_{Z}}Z_{n,j}\hat{D}^{\textsf{in}}_{j,P}\Big).
\end{aligned}
\end{equation}

By defining the quadrature vectors $\hat{\mathbf{X}}=(\hat{x}_1,\hat{x}_2,\ldots,\hat{x}_N)^\top$ and $\hat{\mathbf{P}}=(\hat{p}_1,\hat{p}_2,\ldots,\hat{p}_N)^\top$, we can convert the Heisenberg-Langevin equations into a compact form:
\begin{equation}
\begin{aligned}
 \begin{pmatrix}
 \dot{ \hat{\mathbf{X}} } \\
  \dot{\hat{\mathbf{P}}}  \\
 \end{pmatrix}=&\begin{pmatrix}
  h^\mathbb{X}+YY^\top-ZZ^\top &0 \\
  0 & h^\mathbb{P}+YY^\top-ZZ^\top  \\
 \end{pmatrix}\begin{pmatrix}
  \hat{\mathbf{X}}  \\
  \hat{\mathbf{P}}  \\
 \end{pmatrix}-i\epsilon\begin{pmatrix}
 [\hat{\mathbf{X}},\hat{V}]\\
 [\hat{\mathbf{P}},\hat{V}]
 \end{pmatrix}-\vec{\beta}-\hat{\Omega}^{\textsf{in}}.
\end{aligned}
\end{equation}
Here, the dynamical matrices $h^\mathbb{X}$ and $h^\mathbb{P}$ are
\begin{equation}
\begin{aligned}
h^\mathbb{X}&=-\frac{\kappa}{2}|1\rangle\langle1|+\sum^{N-1}_{n=1}\Big(Je^{A}|n+1\rangle\langle n|-Je^{-A}|n\rangle\langle n+1|\Big),\\
h^\mathbb{P}&=-\frac{\kappa}{2}|1\rangle\langle1|+\sum^{N-1}_{n=1}\Big(Je^{-A}|n+1\rangle\langle n|-Je^{A}|n\rangle\langle n+1|\Big),
\end{aligned}
\end{equation}
the commutation with $\hat{V}$ is
\begin{equation}
\begin{pmatrix}
  [\hat{\mathbf{X}},\hat{V}] \\
  [\hat{\mathbf{P}},\hat{V}]
\end{pmatrix}=([\hat{x}_1,\hat{V}],\ldots,[\hat{x}_N,\hat{V}],[\hat{p}_1,\hat{V}],\ldots,[\hat{p}_N,\hat{V}])^\top,
\end{equation}
the coherent input vector $\vec{\beta}$ is
\begin{equation}
\vec{\beta}=(\sqrt{2\kappa}\beta,0,0,\ldots,0)^\top,\\
\end{equation}
and the quantum noise vectors $\hat{\Omega}^{\textsf{in}}$ are
\begin{equation}
\begin{aligned}
\hat{\Omega}^{\textsf{in}}_{i}&=\sqrt{\kappa}\hat{X}^{\textsf{in}}\delta_{i,1}+\sqrt{2}~\Bigg(\sum_{j=1}^{N_Y}Y_{i,j}\hat{C}_{j,X}^{\textsf{in}}+\sum_{j=1}^{N_Z}Z_{i,j}\hat{D}_{j,X}^{\textsf{in}}\Bigg),\\
\hat{\Omega}^{\textsf{in}}_{i+N}&=\sqrt{\kappa}\hat{P}^{\textsf{in}}\delta_{i,1}+\sqrt{2}~\Bigg(-\sum_{j=1}^{N_Y}Y_{i,j}\hat{C}_{j,P}^{\textsf{in}}+\sum_{j=1}^{N_Z}Z_{i,j}\hat{D}_{j,P}^{\textsf{in}}\Bigg),
\end{aligned}
\end{equation}
for $i=1, \cdots, N$.

\section{Derivations of the SNR per photon}\label{appendixB}

In this appendix, we calculate the signal power, noise power, and the total average photon number when $\epsilon$ is infinitesimal.

According to the Heisenberg-Langevin equations and the definition of  the perturbation Hamiltonian $\epsilon\hat{V}=\epsilon\hat{a}^\dag_N\hat{a}_N$, we have
\begin{equation}
\begin{aligned}
\frac{d\hat{x}_n}{dt}=&~Je^A\hat{x}_{n-1}-Je^{-A}\hat{x}_{n+1}+\epsilon\hat{p}_N\delta_{n,N}-\sqrt{2\kappa}\beta\delta_{n,1}-\frac{\kappa}{2}\hat{x}_1\delta_{n,1}\\
&+\sum_{j=1}^{N_{Y}}\sum_{i=1}^{N}Y_{n,j}Y_{i,j}\hat{x}_i-\sum_{j=1}^{N_{Z}}\sum_{i=1}^{N}Z_{n,j}Z_{i,j}\hat{x}_i\\
&-\sqrt{2}~\Bigg(\sum_{j=1}^{N_{Y}}Y_{n,j}\hat{C}_{j,X}^{\textsf{in}}+\sum_{j=1}^{N_{Z}}Z_{n,j}\hat{D}_{j,X}^{\textsf{in}}\Bigg)-\sqrt{\kappa}\hat{X}^{\textsf{in}}\delta_{n,1},\\
\frac{d\hat{p}_n}{dt}=&~Je^{-A}\hat{p}_{n-1}-Je^{A}\hat{p}_{n+1}-\epsilon\hat{x}_N\delta_{n,N}-\frac{\kappa}{2}\hat{p}_1\delta_{n,1}\\
&+\sum_{j=1}^{N_{Y}}\sum_{i=1}^{N}Y_{n,j}Y_{i,j}\hat{p}_i-\sum_{j=1}^{N_{Z}}\sum_{i=1}^{N}Z_{n,j}Z_{i,j}\hat{p}_i\\
&-\sqrt{2}~\Bigg(-\sum_{j=1}^{N_{Y}}Y_{n,j}\hat{C}_{j,P}^{\textsf{in}}+\sum_{j=1}^{N_{Z}}Z_{n,j}\hat{D}_{j,P}^{\textsf{in}}\Bigg)-\sqrt{\kappa}\hat{P}^{\textsf{in}}\delta_{n,1}.
\end{aligned}
\end{equation}
Since the system is stable, for sufficiently long time, $\hat{x}_n$ and $\hat{p}_n$ can be described as
\begin{equation}\label{appendixC2}
\begin{aligned}
\hat{x}_n=&~\mathbb{H}[\epsilon]^{-1}_{n,1}(\sqrt{2\kappa}\beta+\sqrt{\kappa}\hat{X}^{\textsf{in}}) +\mathbb{H}[\epsilon]^{-1}_{n,N+1}\sqrt{\kappa}\hat{P}^{\textsf{in}}\\
&+\sqrt{2}\sum^N_{i=1}\mathbb{H}[\epsilon]^{-1}_{n,i}\Big(\sum^{N_Y}_{j=1}Y_{i,j}\hat{C}^{\textsf{in}}_{j,X}
+\sum^{N_Z}_{j=1}Z_{i,j}\hat{D}^{\textsf{in}}_{j,X}\Big)\\
&+\sqrt{2}\sum^N_{i=1}\mathbb{H}[\epsilon]^{-1}_{n,N+i}\Big(-\sum^{N_Y}_{j=1}Y_{i,j}\hat{C}^{\textsf{in}}_{j,P}
+\sum^{N_Z}_{j=1}Z_{i,j}\hat{D}^{\textsf{in}}_{j,P}\Big),\\
\hat{p}_n=&~\mathbb{H}[\epsilon]^{-1}_{N+n,1}(\sqrt{2\kappa}\beta+\sqrt{\kappa}\hat{X}^{\textsf{in}}) +\mathbb{H}[\epsilon]^{-1}_{N+n,N+1}\sqrt{\kappa}\hat{P}^{\textsf{in}}\\
&+\sqrt{2}\sum^N_{i=1}\mathbb{H}[\epsilon]^{-1}_{N+n,i}\Big(\sum^{N_Y}_{j=1}
Y_{i,j}\hat{C}^{\textsf{in}}_{j,X}+\sum^{N_Z}_{j=1}Z_{i,j}\hat{D}^{\textsf{in}}_{j,X}\Big)\\
&+\sqrt{2}\sum^N_{i=1}\mathbb{H}[\epsilon]^{-1}_{N+n,N+i}\Big(-\sum^{N_Y}_{j=1}
Y_{i,j}\hat{C}^{\textsf{in}}_{j,P}+\sum^{N_Z}_{j=1}Z_{i,j}\hat{D}^{\textsf{in}}_{j,P}\Big),
\end{aligned}
\end{equation}
where $\mathbb{H}[\epsilon]=\mathbb{H}_1[0]+\mathbb{H}_N[\epsilon]$,
$
\mathbb{H}_1[0]=\begin{pmatrix}
  (Q^{\mathbb{X}})^{-1} &0 \\
  0 & (Q^{\mathbb{P}})^{-1} \\
 \end{pmatrix},
$
$Q^\mathbb{X}=(h^\mathbb{X}+YY^\top-ZZ^\top)^{-1}$, $Q^\mathbb{P}=(h^\mathbb{P}+YY^\top-ZZ^\top)^{-1}$
and $\mathbb{H}_N[\epsilon]=\epsilon |N\rangle\langle2N|-\epsilon|2N\rangle\langle N|$.

Using Dyson's equation and keeping it up to the first order in $\epsilon$, we have
\begin{equation}
\begin{aligned}
\mathbb{H}[\epsilon]^{-1}=(\mathbb{H}_1[0] +\mathbb{H}_N[\epsilon])^{-1} = \mathbb{H}_1[0]^{-1}-\mathbb{H}_1[0]^{-1}\mathbb{H}_N
[\epsilon]\mathbb{H}_1[0]^{-1}.
\end{aligned}
\end{equation}
The detailed calculation of the elements of $\mathbb{H}[\epsilon]^{-1}$ can be found in Appendix~\ref{appendixG}.

It can be computed that the steady state of mode 1 is
\begin{equation}\label{x1p1}
\begin{aligned}
\hat{x}_1=&~Q^{\mathbb{X}}_{1,1}(\sqrt{2\kappa}\beta+\sqrt{\kappa}\hat{X}^{\textsf{in}}) -\epsilon Q^{\mathbb{X}}_{1,N}Q^{\mathbb{P}}_{N,1}\sqrt{\kappa}\hat{P}^{\textsf{in}}\\
&+\sqrt{2}\sum^N_{i=1}Q^{\mathbb{X}}_{1,i}\Big(\sum^{N_Y}_{j=1}Y_{i,j}\hat{C}^{\textsf{in}}_{j,X}+\sum^{N_Z}_{j=1}Z_{i,j}\hat{D}^{\textsf{in}}_{j,X}\Big)\\&-\sqrt{2}\epsilon\sum^N_{i=1}Q^{\mathbb{X}}_{1,N}Q^{\mathbb{P}}_{N,i}\Big(-\sum^{N_Y}_{j=1}Y_{i,j}\hat{C}^{\textsf{in}}_{j,P}+\sum^{N_Z}_{j=1}Z_{i,j}\hat{D}^{\textsf{in}}_{j,P}\Big),\\
\hat{p}_1=&~\epsilon Q^{\mathbb{X}}_{N,1}Q^{\mathbb{P}}_{1,N}(\sqrt{2\kappa}\beta+\sqrt{\kappa}\hat{X}^{\textsf{in}}) +Q^{\mathbb{P}}_{1,1}\sqrt{\kappa}\hat{P}^{\textsf{in}}\\
&+\sqrt{2}\epsilon\sum^N_{i=1}Q^{\mathbb{X}}_{N,i}Q^{\mathbb{P}}_{1,N}\Big(\sum^{N_Y}_{j=1}Y_{i,j}\hat{C}^{\textsf{in}}_{j,X}+\sum^{N_Z}_{j=1}Z_{i,j}\hat{D}^{\textsf{in}}_{j,X}\Big)\\&+\sqrt{2}\sum^N_{i=1}Q^{\mathbb{P}}_{1,i}\Big(-\sum^{N_Y}_{j=1}Y_{i,j}\hat{C}^{\textsf{in}}_{j,P}+\sum^{N_Z}_{j=1}Z_{i,j}\hat{D}^{\textsf{in}}_{j,P}\Big),
\end{aligned}
\end{equation}
where we have used $\mathbb{H}[0]^{-1}=\begin{pmatrix}
  Q^{\mathbb{X}} &0 \\
  0 & Q^{\mathbb{P}}  \\
 \end{pmatrix}$.

From the definition of  $\hat{\mathcal{M}}$, we have
\begin{equation}
\begin{aligned}
\langle\hat{\mathcal{M}}\rangle_{\epsilon}&=\frac{1}{\sqrt{2}i}\Big{(}\langle\hat{\mathcal{B}}\rangle_{\epsilon}-\langle\hat{\mathcal{B}}^\dagger\rangle_{\epsilon}\Big{)}\\
&=\frac{1}{\sqrt{2\tau}i}\int_{0}^{\tau}\sqrt{\kappa}~\Big(\langle\hat{a}_1\rangle_{\epsilon}-\langle\hat{a}_1^\dagger\rangle_{\epsilon}\Big)\;dt\\
&=\frac{1}{\sqrt{2\tau}i}\int_{0}^{\tau}\sqrt{\kappa}\sqrt{2}i\langle\hat{p}_1\rangle_\epsilon \;dt\\
&=\frac{1}{\sqrt{2\tau}i}\int_{0}^{\tau}\sqrt{\kappa}\sqrt{2}i\epsilon Q^{\mathbb{X}}_{N,1}Q^{\mathbb{P}}_{1,N}\sqrt{2\kappa}\beta \;dt\\
&=\sqrt{2\kappa\tau}\epsilon\sqrt{\kappa}\beta Q^{\mathbb{X}}_{N,1}Q^{\mathbb{P}}_{1,N}.
\end{aligned}
\end{equation}

According to the definition of the signal power, we have
\begin{equation}\label{App2signal}
\begin{aligned}
\mathcal{S}(\epsilon)&=\Big|\langle\hat{\mathcal{M}}\rangle_\epsilon-\langle\hat{\mathcal{M}}\rangle_0\Big|^2=
2\epsilon^2\kappa^2\beta^2\tau\Big{|}Q^\mathbb{X}_{N,1}\Big{|}^2\Big{|}Q^\mathbb{P}_{1, N}\Big{|}^2.
\end{aligned}
\end{equation}

For the noise power, recall that only the zeroth order in $\epsilon$ is related to the SNR. Thus,
\begin{equation}
\begin{aligned}
&\hat{\mathcal{M}}\Big|_{\epsilon=0}-\langle\hat{\mathcal{M}}\rangle\Big|_{\epsilon=0}\\ =&\frac{1}{\sqrt{2\tau}i}\int^\tau_0\big{(}\hat{B}^{\textsf{in}}+\sqrt{\kappa}(\hat{a}_1-\langle\hat{a}_1\rangle)
\big|_{\epsilon=0}\big{)}-\big{(}\hat{B}^{\textsf{in}\dagger}+\sqrt{\kappa}(\hat{a}^\dagger_1
-\langle\hat{a}^\dagger_1\rangle)\big|_{\epsilon=0}\big{)}dt\\ =&\frac{1}{\sqrt{2\tau}i}\int^\tau_0 \Bigg[\Big(\frac{1}{\sqrt{2}}+\frac{\kappa}{\sqrt{2}}Q^\mathbb{X}_{1,1}\Big)\hat{X}^{\textsf{in}}+\sqrt{\kappa}\sum^N_{i=1}Q^\mathbb{X}_{1,i}\Big(\sum^{N_Y}_{j=1}Y_{i,j}\hat{C}^{\textsf{in}}_{j,X}+\sum^{N_Z}_{j=1}Z_{i,j}\hat{D}^{\textsf{in}}_{j,X}\Big)\\
&~~~~~~+i\Big(\frac{1}{\sqrt{2}}+\frac{\kappa}{\sqrt{2}}Q^\mathbb{P}_{1,1}\Big)\hat{P}^{\textsf{in}}+i\sqrt{\kappa}\sum^N_{i=1}Q^\mathbb{P}_{1,i}\Big(\sum^{N_Y}_{j=1}Y_{i,j}\hat{C}^{\textsf{in}}_{j,P}+\sum^{N_Z}_{j=1}Z_{i,j}\hat{D}^{\textsf{in}}_{j,P}\Big)\Bigg]\\
&~~~~~~-\Bigg[\Big(\frac{1}{\sqrt{2}}+\frac{\kappa}{\sqrt{2}}Q^\mathbb{X}_{1,1}\Big)\hat{X}^{\textsf{in}}+\sqrt{\kappa}\sum^N_{i=1}Q^\mathbb{X}_{1,i}\Big(\sum^{N_Y}_{j=1}Y_{i,j}\hat{C}^{\textsf{in}}_{j,X}+\sum^{N_Z}_{j=1}Z_{i,j}\hat{D}^{\textsf{in}}_{j,X}\Big)\\
&~~~~~~-i\Big(\frac{1}{\sqrt{2}}+\frac{\kappa}{\sqrt{2}}Q^\mathbb{P}_{1,1}\Big)\hat{P}^{\textsf{in}}-i\sqrt{\kappa}\sum^N_{i=1}Q^\mathbb{P}_{1,i}\Big(\sum^{N_Y}_{j=1}Y_{i,j}\hat{C}^{\textsf{in}}_{j,P}+\sum^{N_Z}_{j=1}Z_{i,j}\hat{D}^{\textsf{in}}_{j,P}\Big)\Bigg]dt\\
=&\frac{1}{\sqrt{2\tau}}\int^\tau_0\sqrt{2}\Big(1+\kappa Q^\mathbb{P}_{1,1}\Big)\hat{P}^{\textsf{in}}+2\sqrt{\kappa}\sum^N_{i=1}Q^\mathbb{P}_{1,i}\Big(\sum^{N_Y}_{j=1}Y_{i,j}\hat{C}^{\textsf{in}}_{j,P}+\sum^{N_Z}_{j=1}Z_{i,j}\hat{D}^{\textsf{in}}_{j,P}\Big)
dt.
\end{aligned}
\end{equation}
Hence, it can be computed that the noise power is
\begin{equation}\label{appendixC7}
\begin{aligned}
\mathcal{N}(0)=\frac{1}{2}(1+\kappa Q_{1,1}^\mathbb{P})^2+\kappa\big{[} Q^\mathbb{P}(YY^\top+ZZ^\top){Q^\mathbb{P}}^{\top}\big{]}_{1,1},
\end{aligned}
\end{equation}
where we have focused on the vacuum noise, i.e., $\bar{n}^{\textsf{th}}_Q=0$.

Following a similar reasoning to the noise power, we are now only concerned about $\bar{n}_{\textsf{tot}}(0)$, the zeroth-order term of the total average photon number with respect to $\epsilon$. In the large-drive limit $|\beta|\geq 1$, we have
\begin{equation}\label{app2ntot}
\begin{aligned}
\bar{n}_{\textsf{tot}}(0)= &\sum_{n=1}^N \langle\hat{a}_n^\dagger\rangle_0\langle\hat{a}_n\rangle_0=\frac{1}{2}\sum_{n=1}^N\langle\hat{x}_n\rangle^2_0+\langle\hat{p}_n\rangle^2_0
=\kappa\beta^2\sum_{n=1}^N\Big|Q^\mathbb{X}_{n,1}\Big|^2
=\kappa\beta^2[{Q^{\mathbb{X}}}^\top Q^\mathbb{X}]_{1,1}.
\end{aligned}
\end{equation}

\section{Stability analysis under $YY^{\top}\neq ZZ^{\top}$}\label{appendixC}

We first briefly introduce the notion  {\it stability} of a linear time-invariant system. 
Consider an $n$-dimensional linear time-invariant system
\begin{equation}\label{linear}
\begin{aligned}
\dot{x}&=Ax,~~~x(0)=x_0;\\
y&=Cx.
\end{aligned}
\end{equation}
The transfer function from the initial condition $x_0$ to the output $y$ reads
\begin{equation}\label{transfer}
G_{x_0\rightarrow y}(s)=C(sI-A)^{-1}x_0,
\end{equation}
whose characteristic equation is given by 
\begin{equation}\label{charactereq}
\det(sI-A)=0.
\end{equation}
The solution to the differential equation (\ref{linear})
can be described as
\begin{equation}\label{output}
y(t)=\sum^{n}_{i=1} K_i \exp\{p_it\},
\end{equation}
where $\{p_i\}$ are the roots of the characteristic equation~(\ref{charactereq}) or the poles of the transfer function~(\ref{transfer}), and $\{K_i\}$ depend on the initial condition and the zero locations of the  transfer function~(\ref{transfer}). Here, we have assumed that all the roots of Eq.~(\ref{charactereq}) are distinct for simplicity. If any poles are repeated, the corresponding coefficient $K_i$ in Eq.~(\ref{output}) must include a polynomial in $t$, but the conclusion is the same.

The system   is {\it stable} if and only if (necessary and sufficient condition) every term in Eq.~(\ref{output}) goes to zero as $t\rightarrow \infty$:
$$\exp\{p_it\}\rightarrow 0 ~\text{for all}~ p_i.$$
This will happen if all the eigenvalues of the system matrix $A$  are strictly in the left half plane, where $\text{Re}\{p_i\} < 0.$ If the system has any poles in the right half plane, then $y(t)\rightarrow \infty$, it is {\it unstable}.
Thus, we can determine the stability of a system by determining whether all  the eigenvalues of the system matrix $A$ sit in the left half plane. A well-known tool to do this is the Routh’s stability criterion (e.g., see Sec. 3.6.3 of [56] of the main text).

We now turn back to our work to employ Routh’s stability criterion to derive the necessary conditions of the elements of $(YY^\top-ZZ^\top)$ to ensure the stability of the system.
\begin{figure}[htbp]
\centering
\subfigure[]{
\includegraphics[scale=0.7]{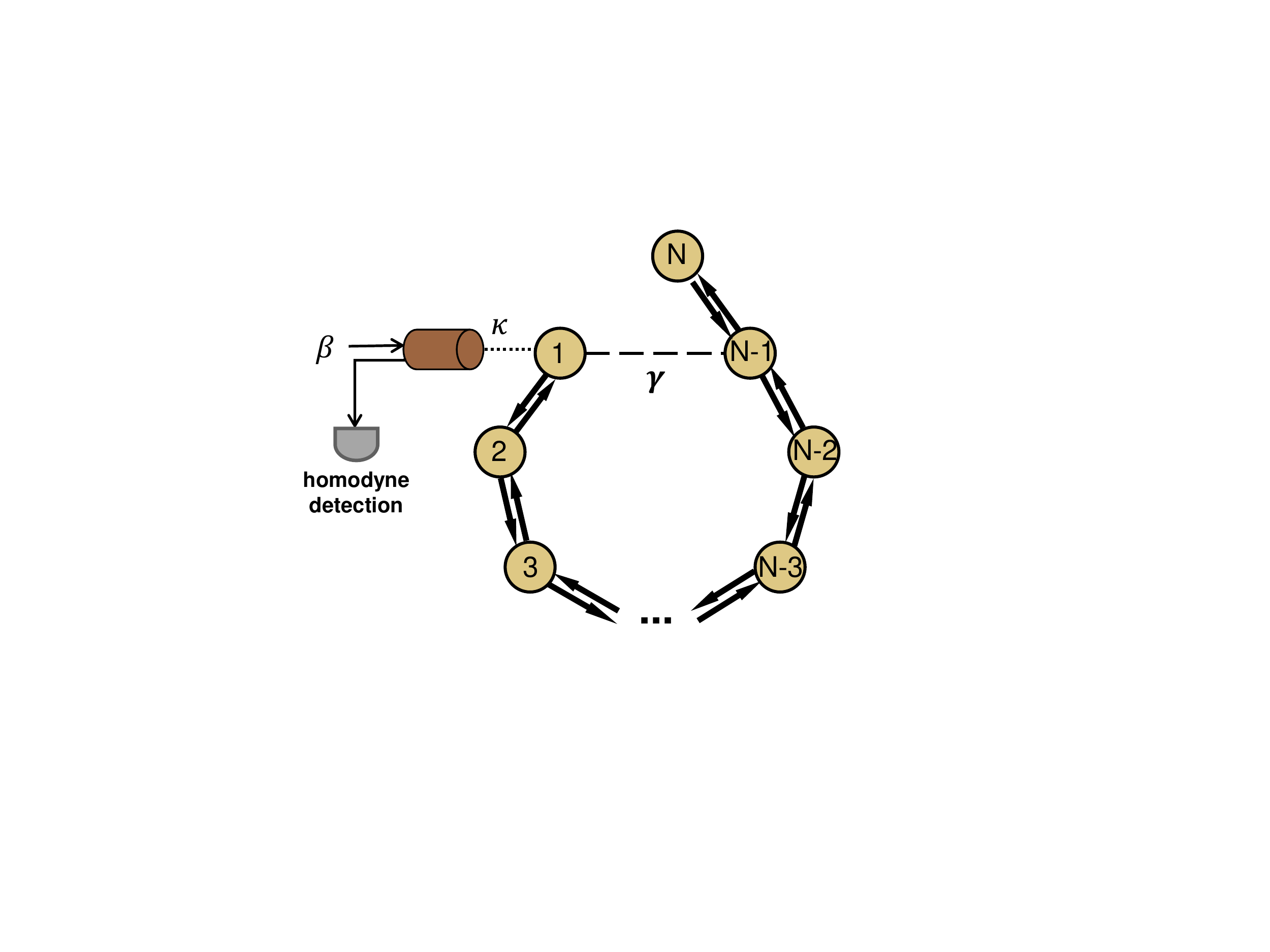}
}
\quad
\subfigure[]{
\includegraphics[scale=0.7]{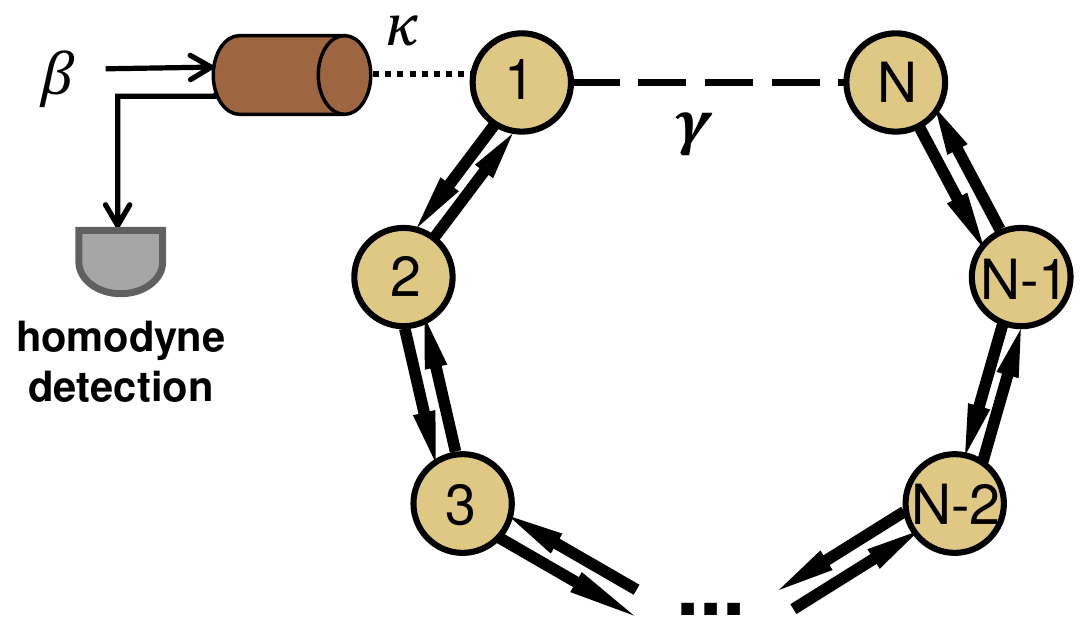}
}
\caption{A schematic of an $N$-site non-Hermitian setup. (a) There is an effective coupling $\gamma$ between the first and the $(N-1)$-th modes.
(b)  There is an effective coupling $\gamma$ between the first and the $N$-th modes.}
\end{figure}

\textbf{Case 1}: $YY^\top-ZZ^\top=\gamma|1\rangle\langle N-1|+\gamma| N-1\rangle\langle1|$.

The noisy setup is illustrated in Fig.~3(a), where there is  an effective coupling $\gamma$ between the first and the $(N-1)$-th modes.
The characteristic polynomial of $(h^\mathbb{X}+YY^\top-ZZ^\top)$  can be calculated in a recursive way as
\begin{equation}\label{characteristic1}
\begin{aligned}
\Big|\lambda I-(h^\mathbb{X}+YY^\top-ZZ^\top)\Big|=&(\lambda+\frac{\kappa}{2})D_{N-1}+J^2D_{N-2}-\gamma^2\lambda D_{N-3}\\
&+\gamma\lambda J^{N-2}e^{-A(N-2)}-\gamma\lambda J^{N-2}e^{A(N-2)},
\end{aligned}
\end{equation}
where
\begin{equation}
\begin{aligned}
D_{N}=\Bigg|\lambda I-\sum^{N-1}_{n=1}\Big(-Je^{-A}|n\rangle\langle n+1|+Je^{A}|n+1\rangle\langle n|\Big)\Bigg|=\sum_{k=0}^{[\frac{N}{2}]}\begin{pmatrix}
N-k\\
k
\end{pmatrix}\lambda^{N-2k}J^{2k},
\end{aligned}
\end{equation}
and $[x]$ is the integer function which returns the largest integer not larger than $x$.

According to Routh's stability criterion, a necessary (but not sufficient) condition for stability is that all the coefficients of $\lambda$  are positive. It can be computed that the coefficient of the first order with respect to $\lambda$  in Eq. \eqref{characteristic1} is
\begin{equation}\label{B3}
\begin{aligned}
&\begin{pmatrix}
\frac{N-1}{2}\\
\frac{N-1}{2}
\end{pmatrix}J^{N-1}+\begin{pmatrix}
\frac{N-1}{2}\\
\frac{N-3}{2}
\end{pmatrix}J^{N-1}-\gamma^2\begin{pmatrix}
\frac{N-3}{2}\\
\frac{N-3}{2}
\end{pmatrix}J^{N-3}+\gamma J^{N-2}e^{-A(N-2)}-\gamma J^{N-2}e^{A(N-2)}\\
&=J^{N-3}\Bigg[\frac{N+1}{2}J^2+\Big(e^{-A(N-2)}-e^{A(N-2)}\Big)J\gamma-\gamma^2\Bigg].
\end{aligned}
\end{equation}
It can be verified that when $e^{-A(N-2)}$ is sufficiently small, to ensure Eq.~(\ref{B3}) be positive, $\gamma$
should satisfy $\gamma_1<\gamma<\gamma_2$, where
\begin{equation}
\begin{aligned}
\gamma_1&=\frac{(e^{-A(N-2)}-e^{A(N-2)})-\sqrt{(e^{-A(N-2)}-e^{A(N-2)})^2+2(N+1)}}{2}J\\ &=-Je^{A(N-2)}+O(e^{-A(N-2)}),\\
\gamma_2&=\frac{(e^{-A(N-2)}-e^{A(N-2)})+\sqrt{(e^{-A(N-2)}-e^{A(N-2)})^2+2(N+1)}}{2}J\\ &=\frac{N+1}{2}Je^{-A(N-2)}+O(e^{-3A(N-2)}).
\end{aligned}
\end{equation}
Thus, if $e^{-A(N-2)}$ is sufficiently small, to ensure the stability of $(h^\mathbb{X}+YY^\top-ZZ^\top)$,  a necessary condition for $\gamma$ is
\begin{equation}\label{l1}
\begin{aligned}
-Je^{A(N-2)}<\gamma<\frac{N+1}{2}Je^{-A(N-2)}.
\end{aligned}
\end{equation}

After a similar analysis with the characteristic polynomial of $(h^\mathbb{P}+YY^\top-ZZ^\top)$, it can be verified that if $e^{-A(N-2)}$ is sufficiently small, to ensure the stability of $(h^\mathbb{P}+YY^\top-ZZ^\top)$, a necessary condition is
\begin{equation}\label{r1}
\begin{aligned}
-\frac{N+1}{2}Je^{-A(N-2)}<\gamma<Je^{A(N-2)}.
\end{aligned}
\end{equation}
Combining Eqs.~(\ref{l1}) and (\ref{r1}), if $YY^\top-ZZ^\top=\gamma|1\rangle\langle N-1|+\gamma| N-1\rangle\langle1|$, a necessary (but not sufficient) condition for the stability of the  system is
\begin{equation}\label{B7}
\begin{aligned}
|\gamma|<\frac{N+1}{2}Je^{-A(N-2)}.
\end{aligned}
\end{equation}

\textbf{Case 2}: $YY^\top-ZZ^\top=\gamma|1\rangle\langle N|+\gamma| N\rangle\langle1|$.

The noisy setup is illustrated in Fig.~3(b), where there is  an effective coupling $\gamma$ between the first and the $N$-th modes. The characteristic polynomial of $(h^\mathbb{X}+YY^\top-ZZ^\top)$ is
\begin{equation}\label{characteristic2}
\begin{aligned}
&\Big|\lambda I-(h^\mathbb{X}+YY^\top-ZZ^\top)\Big|\\=&(\lambda+\frac{\kappa}{2})D_{N-1}+J^2D_{N-2}-\gamma^2 D_{N-2}-\gamma J^{N-1}e^{-A(N-1)}-\gamma J^{N-1}e^{A(N-1)}\\
=&\lambda^N+c_{N-1}\lambda^{N-1}+c_{N-2}\lambda^{N-2}+\ldots+c_1\lambda+c_0,
\end{aligned}
\end{equation}
where
\begin{equation}
\begin{aligned}
c_0=\frac{\kappa}{2}J^{N-1}-\gamma J^{N-1}(e^{-A(N-1)}+e^{A(N-1)}).
\end{aligned}
\end{equation}
According to Routh's stability criterion, a necessary condition for $(h^\mathbb{X}+YY^\top-ZZ^\top)$ to be stable is that all the coefficients of $\lambda$  of the characteristic polynomial  are positive. Thus, from $c_0>0$, we have
\begin{equation}\label{r2}
\begin{aligned}
\gamma<\frac{\kappa}{2}\frac{1}{e^{-A(N-1)}+e^{A(N-1)}}.
\end{aligned}
\end{equation}
%noting that $\frac{\frac{\kappa}{2}}{e^{-A(N-1)}+e^{A(N-1)}}=\frac{\kappa}{2}e^{-A(N-1)}+O(e^{-3A(N-1)})$.
Note that a system is stable if and only if all the elements in the first column of the Routh array are positive. It can be verified that calculating to the fourth row of the Routh array implies that
\begin{equation}\label{l2}
\gamma>-\kappa e^{-A(N-1)}.
\end{equation}
Combining Eqs.~(\ref{r2}) and (\ref{l2}), we obtain that if $YY^\top-ZZ^\top=\gamma|1\rangle\langle N|+\gamma| N\rangle\langle1|$, a necessary (but not sufficient) condition for the  stability of the system is
\begin{equation}\label{B12}
\begin{aligned}
|\gamma|<\kappa e^{-A(N-1)}.
\end{aligned}
\end{equation}

From Eqs.~(\ref{B7}) and (\ref{B12}), it can be seen that if the gain and loss are unbalanced, then  to ensure a stable non-Hermitian sensor, the $(1, N-1)$th and $(1, N)$th elements of the net noise matrix $(YY^\top-ZZ^\top)$ should be exponentially small in terms of the product of $A$ and $N$.

\section{The SNR under condition (\textbf{C1}) }\label{appendixD}

In this appendix, we demonstrate that when the coupling between the system and the gain bath satisfies $Y=0$, the loss structure $Z$ satisfies condition (\textbf{C1}), and the sensing dynamics is stable, the signal power $\mathcal{S}$, noise power $\mathcal{N}$, and the total average photon number $\bar{n}_{\textsf{tot}}(0)$ are the same as those in the ideal noise-free case.

First of all, we prove $(h^\mathbb{P}-ZZ^\top)^{-1}_{1,j}=(h^\mathbb{P})^{-1}_{1,j}$ and $(h^\mathbb{X}-ZZ^\top)^{-1}_{j,1}=(h^\mathbb{X})^{-1}_{j,1}$, for $j=1,\cdots,N$. Indeed, according to the matrix inverse formula,
\begin{equation}
\begin{aligned}
(h^\mathbb{P}-ZZ^\top)^{-1}
&=(h^\mathbb{P}-h^\mathbb{P}CC^\top{ h^{\mathbb{P}}}^{\top})^{-1}\\
&=(h^\mathbb{P})^{-1}+(h^\mathbb{P})^{-1}h^\mathbb{P}CC^\top(-{ h^{\mathbb{P}}}^{\top}(h^{\mathbb{P}})^{-1}
h^\mathbb{P}CC^\top+I)^{-1}{ h^{\mathbb{P}}}^{\top}(h^\mathbb{P})^{-1}\\
&=(h^\mathbb{P})^{-1}+CC^\top(I-{ h^{\mathbb{P}}}^{\top}CC^\top)^{-1}{ h^{\mathbb{P}}}^{\top}(h^\mathbb{P})^{-1}.
\end{aligned}
\end{equation}
Thus, for $j=1,\cdots,N$, we have
\begin{equation}
(h^\mathbb{P}-ZZ^\top)^{-1}_{1,j}
=(h^\mathbb{P})^{-1}_{1,j}+\sum_{k=1}^{N}(CC^\top)_{1,k}[(I-h^{\mathbb{P}\top}CC^\top)^{-1}
{ h^{\mathbb{P}}}^{\top}(h^\mathbb{P})^{-1}]_{k,j}=(h^\mathbb{P})^{-1}_{1,j}.
\end{equation}

Note that since the span of the second through  the $N$-th column of $h^\mathbb{P}$ is the same  as that of ${h^{\mathbb{X}}}^{\top}$, there exists $D\in R^{N_Z\times N}$ with $D_{j,1}=0$ for $j=1,\cdots,N_Z$, such that $Z={h^{\mathbb{X}}}^{\top} D^\top$. Then according to the matrix inverse formula,
\begin{equation}
\begin{aligned}
(h^\mathbb{X}-ZZ^\top)^{-1}
&=(h^\mathbb{X}-{h^{\mathbb{X}}}^{\top}D^\top D h^{\mathbb{X}})^{-1}\\
&=(h^\mathbb{X})^{-1}+(h^\mathbb{X})^{-1}{h^{\mathbb{X}}}^{\top}(I-D^\top Dh^{\mathbb{X}}(h^{\mathbb{X}})^{-1}{h^{\mathbb{X}}}^{\top})^{-1}D^\top D h^{\mathbb{X}}(h^\mathbb{X})^{-1}\\
&=(h^\mathbb{X})^{-1}+(h^\mathbb{X})^{-1}{h^{\mathbb{X}}}^{\top}(I-D^\top D{h^{\mathbb{X}}}^{\top})^{-1}D^\top D.
\end{aligned}
\end{equation}
Thus, for $j=1, \cdots,N$, we have
\begin{equation}
\begin{aligned}
(h^\mathbb{X}-ZZ^\top)^{-1}_{j,1}
&=(h^\mathbb{X})^{-1}_{j,1}+\sum_{k=1}^{N}[(h^\mathbb{X})^{-1}{h^{\mathbb{X}}}^{\top}
(I-D^\top D{h^{\mathbb{X}}}^{\top})^{-1}]_{j,k}(D^\top D)_{k,1}=(h^\mathbb{X})^{-1}_{j,1}.
\end{aligned}
\end{equation}

Therefore, specifically, we have $(h^\mathbb{P}-ZZ^\top)^{-1}_{1,N}=(h^\mathbb{P})^{-1}_{1,N}$ and $(h^\mathbb{X}-ZZ^\top)^{-1}_{N,1}=(h^\mathbb{X})^{-1}_{N,1}$.
Recall that the signal power is in the form of Eq.~\eqref{App2signal}. Thus, we have proved that under $Y=0$, (\textbf{C1}), and if the system is stable,
the signal power of the noisy sensor  is the same as that of the ideal noise-free case.

As for the noise power, according to  Eq.~\eqref{appendixC7}, we only need to prove  $$[(h^\mathbb{P}-ZZ^\top)^{-1}ZZ^\top(h^\mathbb{P}-ZZ^\top)^{-1\top}]_{1,1}=0.$$ In fact,
\begin{equation}
\begin{aligned}
&~~[(h^\mathbb{P}-ZZ^\top)^{-1}ZZ^\top(h^\mathbb{P}-ZZ^\top)^{-1\top}]_{1,1}\\
=&\sum_{k=1}^{N_Z}[(h^\mathbb{P}-ZZ^\top)^{-1}Z]_{1,k}\cdot[(h^\mathbb{P}-ZZ^\top)^{-1}Z]_{1,k}\\
=&\sum_{k=1}^{N_Z}\bigg[\sum_{m=1}^{N}(h^\mathbb{P}-ZZ^\top)^{-1}_{1,m}Z_{m,k}\bigg]\cdot\bigg[\sum_{m=1}^{N}(h^\mathbb{P}-ZZ^\top)^{-1}_{1,m}Z_{m,k}\bigg]\\
=&\sum_{k=1}^{N_Z}\bigg[\sum_{m=1}^{N}(h^\mathbb{P})^{-1}_{1,m}Z_{m,k}\bigg]\cdot\bigg[\sum_{m=1}^{N}(h^\mathbb{P})^{-1}_{1,m}Z_{m,k}\bigg]\\
=&[(h^\mathbb{P})^{-1}ZZ^\top(h^\mathbb{P})^{-1\top}]_{1,1}\\
=&(CC^\top)_{1,1}=0.
\end{aligned}
\end{equation}
Thus, under $Y=0$, (\textbf{C1}), and if the system is stable,
the  noise power is the same as that of the ideal noise-free case.

As for  the total average photon number $\bar{n}_{\textsf{tot}}(0)$, since $Q^{\mathbb{X}}_{j,1}=(h^\mathbb{X}-ZZ^\top)^{-1}_{j,1}=(h^\mathbb{X})^{-1}_{j,1}$, along with Eq.~\eqref{app2ntot}, it is clear that  $\bar{n}_{\textsf{tot}}$ is also the same as that of the ideal noise-free case.

\section{The SNR per photon under conditions \text{(\textbf{C1})} and \text{(\textbf{C2})}}\label{appendixE}

Now we demonstrate that under conditions (\textbf{C1}) and (\textbf{C2}), an exponential enhancement of noisy non-Hermitian sensing  can be revived, that is $\overline{\textrm{SNR}} \propto \exp\{2A(N-1)\}$.

Under  (\textbf{C1}) and (\textbf{C2}), the sensing dynamics is stable.  From  Eqs.~\eqref{App2signal} and \eqref{app2ntot}, it is clear that
\begin{equation}
\begin{aligned}
\mathcal{S}(\epsilon)&=2\epsilon^2\kappa^2\beta^2\tau\Big{|}(h^\mathbb{X})^{-1}_{N,1}\Big{|}^2\Big{|}(h^\mathbb{P})^{-1}_{1, N}\Big{|}^2
=2\epsilon^2\kappa^2\beta^2\tau\Big{(}\frac{2}{\kappa}\Big{)}^4\exp\{4A(N-1)\},\\
\bar{n}_{\textsf{tot}}(0)&=\kappa\beta^2[(h^{\mathbb{X}})^{-1\top} (h^\mathbb{X})^{-1}]_{1,1},
\end{aligned}
\end{equation}
where we have used $(h^\mathbb{X})^{-1}_{N,1}=-\frac{2}{\kappa} \exp\{A(N-1)\}$, and $(h^\mathbb{P})^{-1}_{1, N}=-\frac{2}{\kappa} \exp\{A(N-1)\}$ (see Appendix~\ref{appendixG}).

As for the noise power,
\begin{equation}
\begin{aligned}
\mathcal{N}(0)&=\frac{1}{2}(1+\kappa Q_{1,1}^\mathbb{P})^2+\kappa\big{[} Q^\mathbb{P}(YY^\top+ZZ^\top){Q^\mathbb{P}}^{\top}\big{]}_{1,1}\\
&=\frac{1}{2}(1+\kappa (h^\mathbb{P})^{-1}_{1,1})^2+2\kappa\big{[} (h^\mathbb{P})^{-1}ZZ^\top{(h^\mathbb{P}})^{-1\top}\big{]}_{1,1}\\
&=\frac{1}{2},
\end{aligned}
\end{equation}
where $(h^\mathbb{P})^{-1}_{1,1}=-\frac{2}{\kappa}$ has been used (see Appendix~\ref{appendixG}).
Thus, we now have  $\overline{\textrm{SNR}} \propto \exp\{2A(N-1)\}$.

\section{Derivations of the SNR per photon beyond linear response}\label{appendixF}
In this appendix, we consider the case where the parameter to be detected, $\epsilon_0$, is not infinitesimally small.
Thus, as opposed to the infinitesimal case, not only the linear response
of $\epsilon_0$, but all orders in $\epsilon_0$ of the output field should be
calculated.

The signal power and the total average photon number can be  straightforwardly  calculated  from Eq.~\eqref{appendixC2}.
From the definition of  $\hat{\mathcal{M}}$, we have
\begin{equation}
\begin{aligned}
\langle\hat{\mathcal{M}}\rangle_{\epsilon_0}&=\frac{1}{\sqrt{2}i}\Big{(}\langle\hat{\mathcal{B}}\rangle_{\epsilon_0}-\langle\hat{\mathcal{B}}^\dagger\rangle_{\epsilon_0}\Big{)}\\
&=\frac{1}{\sqrt{2\tau}i}\int_{0}^{\tau}\sqrt{\kappa}\sqrt{2}i\langle\hat{p}_1\rangle_{\epsilon_0} dt\\
&=\frac{1}{\sqrt{2\tau}i}\int_{0}^{\tau}\sqrt{\kappa}\sqrt{2}i\mathbb{H}[\epsilon_0]^{-1}_{N+1,1}\sqrt{2\kappa}\beta dt\\
&=\frac{1}{\sqrt{\tau}}\int_{0}^{\tau}\sqrt{\kappa}\mathbb{H}[\epsilon_0]^{-1}_{N+1,1}\sqrt{2\kappa}\beta dt\\
&=\sqrt{2\kappa\tau}\sqrt{\kappa}\beta\mathbb{H}[\epsilon_0]^{-1}_{N+1,1}.
\end{aligned}
\end{equation}
According to the definition of the signal powers, we have
\begin{equation}
\begin{aligned}
\mathcal{S}(\epsilon_0)&=2\tau\kappa^2\beta\Big|\mathbb{H}[\epsilon_0]^{-1}_{N+1,1}-\mathbb{H}[0]^{-1}_{N+1,1}\Big|^2.
\end{aligned}
\end{equation}
It is clear that under (\textbf{C2}),  $\mathbb{H}[\epsilon_0]$ and  $\mathbb{H}[0]$ are the same as those of the  noise-free case; so is the
signal power accordingly.

The total average photon number $\bar{n}_{\textsf{tot}}$ can be calculated as
\begin{equation}
\begin{aligned}
\bar{n}_{\textsf{tot}}&=\frac{\bar{n}_{\textsf{tot}}(0)+\bar{n}_{\textsf{tot}}(\epsilon_0)}{2}\\
&=\frac{\sum_{n=1}^N\langle\hat{x}_n\rangle^2_0+\langle\hat{p}_n\rangle^2_0}{4}+\frac{\sum_{n=1}^N\langle\hat{x}_n\rangle^2_{\epsilon_0}+\langle\hat{p}_n\rangle^2_{\epsilon_0}}{4}\\
&=\frac{\kappa\beta^2\sum_{n=1}^{N}|(h^{\mathbb{X}})^{-1}_{n,1}\big|^2}{2}+\frac{\kappa\beta^2\sum_{n=1}^{N}\Big(\big|\mathbb{H}[\epsilon_0]^{-1}_{n,1}\big|^2+\big|\mathbb{H}[\epsilon_0]^{-1}_{N+n,1}\big|^2\Big)}{2}.
\end{aligned}
\end{equation}
Similar to the signal power, we can see that under (\textbf{C2}), the total average photon number is the same as that of the noise-free case.

We now calculate the noise power. Since $\epsilon_0$ is no longer infinitesimal, we have to compute all the orders of the output field with respect to $\epsilon_0$. According to the definition of $\hat{\mathcal{M}}$, we have
\begin{equation}
\begin{aligned}
&\hat{\mathcal{M}}\Big|_{\epsilon=\epsilon_0}-\langle\hat{\mathcal{M}}\rangle\Big|_{\epsilon=\epsilon_0}\\
=&\frac{1}{\sqrt{\tau}}\int^{\tau}_{0}\Bigg{\{}\kappa\mathbb{H}[\epsilon_0]^{-1}_{N+1,1}\hat{X}^{\textsf{in}}+\kappa\mathbb{H}[\epsilon_0]^{-1}_{N+1,N+1}\hat{P}^{\textsf{in}}+\hat{P}^{\textsf{in}}+\sum_{i=1}^{N}\mathbb{H}[\epsilon_0]^{-1}_{N+1,i}\sqrt{2\kappa}~\Big(\sum_{j=1}^{N_Y}Y_{i,j}\hat{C}^{\textsf{in}}_{j,X}+\sum_{j=1}^{N_Z}Z_{i,j}\hat{D}^{\textsf{in}}_{j,X}\Big)\\
&~~~~~+\sum_{i=1}^{N}\mathbb{H}[\epsilon_0]^{-1}_{N+1,N+i}\sqrt{2\kappa}~\Big(-\sum_{j=1}^{N_Y}Y_{i,j}\hat{C}^{\textsf{in}}_{j,P}+\sum_{j=1}^{N_Z}Z_{i,j}\hat{D}^{\textsf{in}}_{j,P}\Big)\Bigg{\}}dt.
\end{aligned}
\end{equation}
Then the noise power $\mathcal{N}(\epsilon_0)$ can be calculated as
\begin{equation}
\begin{aligned}
\mathcal{N}(\epsilon_0)&=\langle\hat{\mathcal{M}}^2\rangle_{\epsilon_0}-\langle\hat{\mathcal{M}}\rangle_{\epsilon_0}^2\\
&=\frac{1}{2}\Bigg\{\kappa^2(\mathbb{H}[\epsilon_0]^{-1}_{N+1,1})^2+(1+\kappa\mathbb{H}[\epsilon_0]^{-1}_{N+1,N+1})^2+2\kappa\Big[\mathbb{H}[\epsilon_0]^{-1}\begin{pmatrix}
                                                                                                                                                                    Y \\
                                                                                                                                                                    0
                                                                                                                                                                  \end{pmatrix}\cdot\begin{pmatrix}
                                                                                                                                                                                      Y^\top & 0
                                                                                                                                                                                    \end{pmatrix}\mathbb{H}[\epsilon_0]^{-1\top}\Big]_{N+1,N+1}\\
&~~~+2\kappa\Big[\mathbb{H}[\epsilon_0]^{-1}\begin{pmatrix}
                                                                                                                                                                    Z \\
                                                                                                                                                                    0
                                                                                                                                                                  \end{pmatrix}\cdot\begin{pmatrix}
                                                                                                                                                                                      Z^\top & 0
                                                                                                                                                                                    \end{pmatrix}\mathbb{H}[\epsilon_0]^{-1\top}\Big]_{N+1,N+1}
+2\kappa\Big[\mathbb{H}[\epsilon_0]^{-1}\begin{pmatrix}
                                                                                                                                                                    0 \\
                                                                                                                                                                    Y
                                                                                                                                                                  \end{pmatrix}\cdot\begin{pmatrix}
                                                                                                                                                                                     0 & Y^\top
                                                                                                                                                                                    \end{pmatrix}\mathbb{H}[\epsilon_0]^{-1\top}\Big]_{N+1,N+1}\\
&~~~+2\kappa\Big[\mathbb{H}[\epsilon_0]^{-1}\begin{pmatrix}
                                                                                                                                                                    0 \\
                                                                                                                                                                    Z
                                                                                                                                                                  \end{pmatrix}\cdot\begin{pmatrix}
                                                                                                                                                                                      0 & Z^\top
                                                                                                                                                                                    \end{pmatrix}\mathbb{H}[\epsilon_0]^{-1\top}\Big]_{N+1,N+1}
\Bigg\}\\
&=\frac{1}{2}\Bigg\{\!\kappa^2(\mathbb{H}[\epsilon_0]^{-1}_{\!N\!+\!1,1})^2\!+\!(1\!+\!\kappa\mathbb{H}[\epsilon_0]^{-1}_{\!N\!+\!1,N\!+\!1})^2\!+\!2\kappa\bigg[\mathbb{H}[\epsilon_0]^{-1}\!\begin{pmatrix}
                                                                                                                                                               YY^\top\!\!+\!ZZ^\top & 0 \\
                                                                                                                                                               0 & YY^\top\!\!+\!ZZ^\top
                                                                                                                                                             \end{pmatrix}\!\mathbb{H}[\epsilon_0]^{-1\top}\bigg]_{\!N\!+\!1,N\!+\!1}
\!\Bigg\}\\
&=\frac{1}{2}\Bigg\{\!\kappa^2(\mathbb{H}[\epsilon_0]^{-1}_{N\!+\!1,1})^2\!+\!(1\!+\!\kappa\mathbb{H}[\epsilon_0]^{-1}_{N\!+\!1,N\!+\!1})^2\!+\!4\kappa\bigg[\mathbb{H}[\epsilon_0]^{-1}\!\begin{pmatrix}
                                                                                                                                                               ZZ^\top & 0 \\
                                                                                                                                                               0 & ZZ^\top
                                                                                                                                                             \end{pmatrix}\!\mathbb{H}[\epsilon_0]^{-1\top}\bigg]_{N\!+\!1,N\!+\!1}
\!\Bigg\}.
\end{aligned}
\end{equation}
Combining with Eq.~\eqref{appendixC7}, the total noise power $\mathcal{N}$ beyond linear response reads
\begin{equation}
\begin{aligned}
\mathcal{N}=&\frac{\mathcal{N}(0)+\mathcal{N}(\epsilon_0)}{2}\\
=&\frac{1}{4}\Bigg\{\kappa^2(\mathbb{H}[\epsilon_0]^{-1}_{N+1,1})^2+(1+\kappa\mathbb{H}[\epsilon_0]^{-1}_{N+1,N+1})^2+(1+\kappa(h^{\mathbb{P}})^{-1}_{1,1})^2+4\kappa\big[(h^{\mathbb{P}})^{-1}ZZ^\top (h^{\mathbb{P}})^{-1\top}\big]_{1,1}\\
&+4\kappa\bigg[\mathbb{H}[\epsilon_0]^{-1}\begin{pmatrix}
                                                                                                                                                                     ZZ^\top & 0 \\
                                                                                                                                                                    0 & ZZ^\top
                                                                                                                                                                   \end{pmatrix}(\mathbb{H}[\epsilon_0]^{-1})^\top\bigg]_{N+1,N+1}\Bigg\}.
\end{aligned}
\end{equation}
Note that in the above equations the first three terms do not depend on the noise.  To regain the SNR per photon in the ideal case, it is necessary to design the loss structure $Z$ such that the total effect of the noise on the noise power $\mathcal{N}$ is canceled out. The last term in the above equation can be expressed as
\begin{equation}
\begin{aligned}
&\bigg[\mathbb{H}[\epsilon_0]^{-1}\begin{pmatrix}
                                                                                                                                                                     ZZ^\top & 0 \\
                                                                                                                                                                    0 & ZZ^\top
                                                                                                                                                                   \end{pmatrix}(\mathbb{H}[\epsilon_0]^{-1})^\top\bigg]_{N+1,N+1}\\
=&\sum_{j=1}^{N_Z}\Bigg(\sum_{i=1}^{N}\mathbb{H}[\epsilon_0]^{-1}_{N+1,i}Z_{i,j}\Bigg)^2+\sum_{j=1}^{N_Z}\Bigg(\sum_{i=1}^{N}\mathbb{H}[\epsilon_0]^{-1}_{N+1,N+i}Z_{i,j}\Bigg)^2\\
=&\sum_{j=1}^{N_Z}\Bigg(\sum_{i=1}^{N}-\frac{2\epsilon }{\kappa} h^{-1}_{N,i}\frac{1}{1+\epsilon^2\frac{4}{\kappa^2}}e^{A(2N-1-i)}Z_{i,j}\Bigg)^2+\sum_{j=1}^{N_Z}\Bigg( \sum_{i=1}^{N}\bigg(h^{-1}_{N,i}(\frac{1}{1+\epsilon^2\frac{4}{\kappa^2}}-1)+h^{-1}_{1,i}\bigg)e^{A(i-1)}Z_{i,j} \Bigg)^2\\
=&\frac{\frac{4\epsilon^2}{\kappa^2}}{1+\epsilon^2\frac{4}{\kappa^2}}e^{2A(N-1)}\sum_{j=1}^{N_Z}\sum^{N}_{i=1}\Bigg(\sum_{i=1}^{N}(h^{\mathbb{X}})^{-1}_{N,i}Z_{i,j}\Bigg)^2
+\Big(\frac{1}{1+\epsilon^2\frac{4}{\kappa^2}}-1\Big)^2e^{2A(N-1)}\sum_{j=1}^{N_Z}\Bigg(\sum_{i=1}^{N}(h^{\mathbb{P}})^{-1}_{N,i}Z_{i,j}\Bigg)^2\\
&+\sum^{N_Z}_{j=1}\Bigg(\sum_{i=1}^{N}(h^{\mathbb{P}})^{-1}_{1,i}Z_{i,j}\Bigg)^2+2\Big(\frac{1}{1+\epsilon^2\frac{4}{\kappa^2}}-1\Big)e^{A(N-1)}\sum_{j=1}^{N_Z}\Bigg(\sum_{i=1}^{N}(h^{\mathbb{P}})^{-1}_{N,i}Z_{i,j}\Bigg)\Bigg(\sum_{i=1}^{N}(h^{\mathbb{P}})^{-1}_{1,i}Z_{i,j}\Bigg)\\
=&\frac{\frac{4\epsilon^2}{\kappa^2}}{(1+\epsilon^2\frac{4}{\kappa^2})^2}e^{2A(N-1)}\Big[(h^{\mathbb{X}})^{-1}ZZ^\top (h^{\mathbb{X}})^{-1\top}\Big]_{N,N}+\Big(\frac{1}{1+\epsilon^2\frac{4}{\kappa^2}}-1\Big)^2e^{2A(N-1)}\Big[(h^{\mathbb{P}})^{-1}ZZ^\top (h^{\mathbb{P}})^{-1\top}\Big]_{N,N}\\
&+\Big[(h^{\mathbb{P}})^{-1}ZZ^\top (h^{\mathbb{P}})^{-1\top}\Big]_{1,1}+2\Big(\frac{1}{1+\epsilon^2\frac{4}{\kappa^2}}-1\Big)e^{A(N-1)}\Big[(h^{\mathbb{P}})^{-1}ZZ^\top (h^{\mathbb{P}})^{-1\top}\Big]_{N,1},
\end{aligned}
\end{equation}
where the details of calculating the elements of  $\mathbb{H}[\epsilon_0]^{-1}$ can be found in Appendix~\ref{appendixH}.

It can be seen that under  (\textbf{C3}) and (\textbf{C4}),
\begin{equation}
\begin{aligned}
\Big[(h^{\mathbb{P}})^{-1}ZZ^\top(h^{\mathbb{P}})^{-1\top}\Big]_{1,1}&=\big[CC^\top\big]_{1,1}=0,\\
\Big[(h^{\mathbb{P}})^{-1}ZZ^\top(h^{\mathbb{P}})^{-1\top}\Big]_{N,N}&=\big[CC^\top\big]_{N,N}=0,\\
\Big[(h^{\mathbb{P}})^{-1}ZZ^\top(h^{\mathbb{P}})^{-1\top}\Big]_{N,1}&=\big[CC^\top\big]_{N,1}=0,\\
\Big[(h^{\mathbb{X}})^{-1}ZZ^\top(h^{\mathbb{X}})^{-1\top}\Big]_{N,N}&=0.
\end{aligned}
\end{equation}
Hence, under (\textbf{C2}), (\textbf{C3}) and (\textbf{C4}), the  noise power is the same as that of the ideal case.  Thus, we have regained the best sensitivity, which is the same as when there is no noise.

\section{Calculation of the elements of  $\mathbb{H}[\epsilon]^{-1}$}\label{appendixG}

In this section, we calculate the elements of $\mathbb{H}[\epsilon]^{-1}$ when $\epsilon$ is infinitesimal. Using Dyson's equation and keeping it up to the first order in $\epsilon$, we have
\begin{equation}\label{HH1HN}
\begin{aligned}
\mathbb{H}[\epsilon]^{-1}=(\mathbb{H}_1[0] +\mathbb{H}_N[\epsilon])^{-1} = \mathbb{H}_1[0]^{-1}-\mathbb{H}_1[0]^{-1}\mathbb{H}_N
[\epsilon]\mathbb{H}_1[0]^{-1},
\end{aligned}
\end{equation}
where $\mathbb{H}_1[0]=\begin{pmatrix}
  h^{\mathbb{X}}+YY^\top-ZZ^\top &0 \\
  0 & h^{\mathbb{P}}+YY^\top-ZZ^\top  \\
 \end{pmatrix}$, and $\mathbb{H}_N
[\epsilon]=\epsilon|N\rangle\langle2N|-\epsilon|2N\rangle\langle N|$.

We only compute $\mathbb{H}[\epsilon]^{-1}_{N+1,1}$ as an illustration, and the other elements can be computed in a similar way.
Multiplying Eq.~\eqref{HH1HN} from the left by $\langle N+1|$ and from the right by $|1\rangle$ yields
\begin{equation}
\begin{aligned}
\mathbb{H}[\epsilon]^{-1}_{N+1,1}&= \mathbb{H}_1[0]^{-1}_{N+1,1}+\epsilon\mathbb{H}_1[0]^{-1}_{N+1,2N}\mathbb{H}_1[0]^{-1}_{N,1}\\
&=\epsilon (h^\mathbb{P}+YY^\top-ZZ^\top)^{-1}_{1,N}(h^\mathbb{X}+YY^\top-ZZ^\top)^{-1}_{N,1}\\
&=\epsilon Q^{\mathbb{X}}_{N,1}Q^{\mathbb{P}}_{1,N}.
\end{aligned}
\end{equation}

Under (\textbf{C2}), we only need to compute the elements of $(h^{\mathbb{X}})^{-1}$ and $(h^{\mathbb{P}})^{-1}$. Defining $T=\text{diag}\{1,e^A,e^{2A},\ldots,e^{A(N-1)}\}$, it can be verified that
\begin{equation}
\begin{aligned}
(h^\mathbb{X})^{-1}=Th^{-1}T^{-1},\\
(h^\mathbb{P})^{-1}=T^{-1}h^{-1}T,
\end{aligned}
\end{equation}
where
\begin{equation}
h=-\frac{\kappa}{2}|1\rangle\langle1|+\sum^{N-1}_{n=1}\Big(-J|n\rangle\langle n+1|+J|n+1\rangle\langle n|\Big).
\end{equation}
Then the elements of $(h^\mathbb{X})^{-1}$ are related to the elements of $h^{-1}$ as
\begin{equation}
\begin{aligned}
(h^\mathbb{X})^{-1}_{i,j}=h^{-1}_{i,j}e^{A(i-j)};
\end{aligned}
\end{equation}
while the elements of $(h^\mathbb{P})^{-1}$ relate to the elements of $h^{-1}$ as
\begin{equation}
\begin{aligned}
(h^\mathbb{P})^{-1}_{i,j}=h^{-1}_{i,j}e^{A(j-i)}.
\end{aligned}
\end{equation}
We now  introduce how to compute the elements of $h^{-1}$. According to the definition of $h$, we have
\begin{equation}
\begin{aligned}
\mathbb{I}=\left(J\sum^{N-1}_{n=1} \left(|n+1\rangle\langle n|-|n \rangle\langle n+1| \right)-\frac{\kappa}{2}|1\rangle\langle 1|\right)h^{-1},
\end{aligned}
\end{equation}
where $\mathbb{I}$ is the $N\times N$ identity matrix. Multiplying this equation from the left by $\langle i|$ and the right by $|1\rangle$, for $i=1, \cdots, N$, yields
\begin{equation}
\begin{aligned}
1=&-J\langle2|h^{-1}|1\rangle-\frac{\kappa}{2}\langle1|h^{-1}|1\rangle,\\
0=&(J\langle1|-J\langle3|)h^{-1}|1\rangle,\\
0=&(J\langle2|-J\langle4|)h^{-1}|1\rangle,\\
\vdots\\
0=&(J\langle N-2|-J\langle N|)h^{-1}|1\rangle,\\
0=&J\langle N-1|h^{-1}|1\rangle.
\end{aligned}
\end{equation}
Simplifying the above recursive formula, we have
\begin{equation}
\begin{aligned}
h^{-1}_{1,1}&=h^{-1}_{3,1}=\cdots=h^{-1}_{N,1}=-\frac{2}{\kappa},\\
h^{-1}_{2,1}&=h^{-1}_{4,1}=\cdots=h^{-1}_{N-1,1}=0.
\end{aligned}
\end{equation}
The other elements can be computed in a similar way.

\section{Calculation of the elements  of $\mathbb{H}[\epsilon_0]^{-1}$ }\label{appendixH}

In this section, we calculate the elements of $\mathbb{H}[\epsilon_0]^{-1}$ under (\textbf{C2}). Since the parameter  $\epsilon_0$ is not infinitesimal, we have to consider all orders of  $\epsilon_0$.

Define
\begin{equation}
\begin{aligned}
\tilde{\mathbb{H}}_1[0]&=\begin{pmatrix}
                           T & 0 \\
                           0 & T^{-1}
                         \end{pmatrix}^{-1}\mathbb{H}_1[0]\begin{pmatrix}
                           T & 0 \\
                           0 & T^{-1}
                         \end{pmatrix}=\begin{pmatrix}
                          h & 0 \\
                          0 & h
                        \end{pmatrix},\\
\tilde{\mathbb{H}}_N[\epsilon_0]&=\begin{pmatrix}
                           T & 0 \\
                           0 & T^{-1}
                         \end{pmatrix}^{-1}\mathbb{H}_N[\epsilon_0]\begin{pmatrix}
                           T & 0 \\
                           0 & T^{-1}
                         \end{pmatrix}=\epsilon_0e^{-2A(N-1)}|N\rangle\langle2N|-\epsilon_0e^{2A(N-1)}|2N\rangle\langle N|.
\end{aligned}
\end{equation}
 It can be seen that \begin{equation}
\begin{aligned}
\big(\mathbb{H}_1[0]+\mathbb{H}_N[\epsilon_0]\big)^{-1}_{i,j}=\big(\tilde{\mathbb{H}}_1[0]+\tilde{\mathbb{H}}_N[\epsilon_0]\big)^{-1}_{i,j}e^{A(i-j)},   ~\text{for}~ i,~j=1,\cdots, N.
\end{aligned}
\end{equation} The other elements have similar relationships. Thus, to calculate the elements of $\mathbb{H}[\epsilon_0]^{-1}=\big(\mathbb{H}_1[0]+\mathbb{H}_N[\epsilon_0]\big)^{-1}$, we can calculate the elements of $\big(\tilde{\mathbb{H}}_1[0]+\tilde{\mathbb{H}}_N[\epsilon_0]\big)^{-1}$. It can be verified that
\begin{equation}
\begin{aligned}
\big(\tilde{\mathbb{H}}_1[0]+\tilde{\mathbb{H}}_N[\epsilon_0]\big)^{-1}=&\sum_{n=1}\Big\{(-1)^n\epsilon_0^{2n}\tilde{\mathbb{H}}_1[0]^{-1}|N\rangle(h^{-1}_{N,N})^{2n-1}\langle N|\tilde{\mathbb{H}}_1[0]^{-1}\\
&+(-1)^n\epsilon_0^{2n}\tilde{\mathbb{H}}_1[0]^{-1}|2N\rangle(h^{-1}_{N,N})^{2n-1}\langle 2N|\tilde{\mathbb{H}}_1[0]^{-1}\Big\}+\tilde{\mathbb{H}}_1[0]^{-1}\\
&+\sum_{n=0}\Big\{(-1)^{n+1}\epsilon_0^{2n+1}\tilde{\mathbb{H}}_1[0]^{-1}|N\rangle(h^{-1}_{N,N})^{2n}\langle 2N|\tilde{\mathbb{H}}_1[0]^{-1}e^{-2A(N-1)}\\
&+(-1)^n\epsilon_0^{2n+1}\tilde{\mathbb{H}}_1[0]^{-1}|2N\rangle(h^{-1}_{N,N})^{2n}\langle N|\tilde{\mathbb{H}}_1[0]^{-1}e^{2A(N-1)}\Big\}.
\end{aligned}
\end{equation}
Multiplying this equation from the left by $\langle 1|$ and the right by $|1\rangle$, yields
\begin{equation}
\begin{aligned}
\big(\tilde{\mathbb{H}}_1[0]+\tilde{\mathbb{H}}_N[\epsilon_0]\big)^{-1}_{1,1}=\sum_{n=1}(-1)^n\epsilon_0^{2n}\Big{(}-\frac{2}{\kappa}\Big{)}^{2n+1}+\Big{(}-\frac{2}{\kappa}\Big{)}=-\frac{2}{\kappa}\cdot\frac{1}{1+\epsilon_0^2\frac{4}{\kappa^2}},
\end{aligned}
\end{equation}
and then \begin{equation}\big(\mathbb{H}_1[0]+\mathbb{H}_N[\epsilon_0]\big)^{-1}_{1,1}=\big(\tilde{\mathbb{H}}_1[0]+\tilde{\mathbb{H}}_N[\epsilon_0]\big)^{-1}_{1,1}e^{A(1-1)}=-\frac{2}{\kappa}\cdot\frac{1}{1+\epsilon_0^2\frac{4}{\kappa^2}}.\end{equation}
The other elements can be computed in a similar way.

\end{widetext}

\end{document}